\documentclass[prd,showpacs,preprint]{revtex4}

\usepackage{amsmath}
\usepackage{amsfonts}
\usepackage{amssymb}
\usepackage{latexsym}
\usepackage[dvips]{graphicx}
\usepackage{graphics}
\usepackage{color}

\begin{document}

\title{Particle--Like Description in Quintessential Cosmology} 
\author{Marek Szyd{\l}owski}
\email{uoszydlo@cyf-kr.edu.pl}
\affiliation{Astronomical Observatory, Jagiellonian University,\\ Orla 171, 30-244 Krak\'ow, Poland}
\author{Wojciech Czaja}
\email{czaja@oa.uj.edu.pl}
\affiliation{Astronomical Observatory, Jagiellonian University,\\ Orla 171, 30-244 Krak\'ow, Poland}
\date{\today}

\begin{abstract}
Assuming  equation of state for quintessential matter: $p = w(z)\rho$, we analyse dynamical behaviour
of the scale factor in FRW cosmologies. It is shown that its dynamics is formally equivalent to that of 
a classical particle under the action of 1D potential $V(a)$. It is shown that Hamiltonian method can be 
easily implemented to obtain a classification of all cosmological solutions in the phase space as well as 
in the configurational space. Examples taken from modern cosmology \cite{kambenalc,corasaniti} illustrate 
the effectiveness of the presented approach. Advantages of representing dynamics as a 1D Hamiltonian flow, 
in the analysis of acceleration and horizon problems, are presented.  The distant supernovae type Ia data 
are used to reconstruct the expansion scenario. The inverse problem of reconstructing the Hamiltonian 
dynamics (i.e. potential function) from the luminosity distance function $d_{L}(z)$ for supernovae is also 
considered.
\end{abstract}


\maketitle

\section{Introduction}

The Newtonian analogue to the Friedmann-Robertson-Walker models (hereafter FRW) was first 
considered by Milne and McCrea who used classical mechanics concepts to describe the expansion of 
the universe \cite{milne,mccrea}. This approach has some difficulties connected with description of 
an infinite homogenous Newtonian cosmology \cite{harrison}. Pressure, in contrast to Newtonian theory, 
has only relativistic character \cite{ellis}. In this way, a quintessential universe has 
no Newtonian analogues. However, pressure seems to be play important role in the evolution of the real 
universe. Recently found accelerated expansion of our universe is due to the presence of a certain vacuum 
energy with the negative pressure equation of state \cite{perlmutter}-\cite{schmidt}. This type of energy is 
called dark energy and equation of state, in the general quintessential form, is usually postulated in the 
analysis of dynamics \cite{caldwell}.

Lima et al. \cite{lima} considered the problem of reduction of FRW cosmologies with the equation of 
state $p \propto \rho$ to the problem of classical particle under 1-dimensional homogenous potential. 
We generalize this description for the FRW quintessential cosmologies. 

Recent observations of supernovae (SNIa) have revealed that this dark energy dominates the evolution
of the universe. The most natural candidate to represent dark energy seems to be the cosmological constant.
However, it is necessary to introduce a fine tunning of 120 orders of magnitude in order to obtain the 
agreement with observations. Another popular idea is the so--called quintessence, a self-interacting scalar 
field \cite{sahni}, but the quintessence program suffers from the fine tunning of microphysical parameters. 
In this work, we discuss the possibility that the dark energy is characterized by the equation of state
\begin{equation}
p = w(a) \rho + 0 = w(a(z)) \rho + 0,
\label{eqstate}
\end{equation}
where $z + 1 = a^{-1}$ is the relation between redshift $z$ and the scale factor $a(t)$ in FRW models.

The main aim of this paper is to reduce the FRW dynamics (with the equation of state in the form 
(\ref{eqstate})) to the form of the Newtonian equation of motion for a unit mass particle in 1D potential. 
Therefore, we extend classical results \cite{milne}-\cite{harrison} to the quintessential cosmologies. 
On the other hand, the question concerning the reconstruction of the equation of motion (the same concerns 
reconstruction form of equation of state \cite{sahni}) from the observational data becomes natural. 
We consider problem of reconstructing the potential function for the particle--like description of dynamics.

Organization of our paper is the following. In section 2, different methods of reducing the FRW 
quintessential cosmology (Q.C.) to the 1D Hamiltonian flow is presented. Some applications of this formalism 
are given in section 3. In section 2, we present simple analysis methods of classical cosmological problems 
such as the horizon problem, acceleration or cosmological constant problems. Finally, in section 4 the 
problem of reconstructing dynamics (potential function) from $d_{L}(z)$ in a model independent manner 
(as far as possible) is investigated.

\section{Hamiltonian dynamics of the Q.C. models}

We consider dark energy description in terms of single $w(a(z))$ but it is always
possible to treat the equation of state of dark energy in terms of total pressure versus total density
ratio. Moreover, in any case our treatment can be generalized to the case of many-component of 
non-interacting fluids.

After assuming the quintessential fluid with the equation of state (\ref{eqstate}), where $p$ is the pressure
and $\rho$ the energy density, the dynamics of homogenous and isotropic models can be described by the 
following set of equations
\begin{equation}
\frac{\ddot{a}}{a} = -\frac{1}{6}(1 + 3 w(a)) \rho + \frac{\Lambda}{3},
\label{dyn1}
\end{equation}
\begin{equation}
\frac{d \rho}{d t} = - 3 \frac{\dot{a}}{a}(1 + w(a)) \rho. 
\label{dyn2}
\end{equation}
Equation (\ref{dyn1}) is the Raychaudhuri equation while eq. (\ref{dyn2}) is the continuity equation for 
fluid with density $\rho$ and equation of state (\ref{eqstate}), $\dot{} \equiv \frac{d}{dt}$, $t$
-- cosmological time.

The first integral of system (\ref{dyn1})-(\ref{dyn2}) is the Friedmann equation
\begin{equation}
\rho = 3 \frac{\dot{a}^{2}}{a^{2}} + 3 \frac{k}{a^{2}} - \Lambda,
\label{friedmann1}
\end{equation}
where $H = d(\ln{a})/dt$ is the Hubble function, $\Lambda$ is the cosmological constant and $k$ is the 
curvature index.

After substitution (\ref{friedmann1}) into (\ref{dyn1}) we obtain
\begin{equation}
\ddot{a} + \psi(a) \dot{a}^{2} + \kappa(a) = 0,
\label{dyn3}
\end{equation}
where
\begin{equation}
\psi(a) \equiv \frac{1 + 3 w(a)}{2a},
\label{psi}
\end{equation}
\begin{equation}
\kappa(a) \equiv \frac{1}{2}\Bigg[(1 + 3 w(a)) \frac{k}{a} - (1 + w(a)) \Lambda a\Bigg].
\label{kappa}
\end{equation}
There are two different methods of reducting (\ref{dyn3}) to the form of the Newtonian equation 
of motion (i.e.\ of elimination of the $\psi(a)\dot{a}^{2}$ type term from (\ref{dyn3})).

\noindent
{\bf 1.} The reparametrization of the time variable 
\begin{equation}
t \rightarrow \tau:\hspace{10mm} dt = \varphi(a(\tau)) d\tau,
\label{reparam1}
\end{equation}
\begin{equation}
\varphi \equiv \exp \int_{}^{a}\psi(a)da
\label{reparam2}
\end{equation}
reduces basic equation (\ref{dyn3}) to
\begin{equation}
a'' \equiv \frac{d^{2}a}{d\tau^{2}} = -\frac{\partial V}{\partial a},
\label{reparam3}
\end{equation}
where
\begin{equation}
V(a) = \int_{}^{a}\kappa(a)\varphi^{2}(a)da
\label{reparam4}
\end{equation}
is the potential function for system (\ref{reparam3}). Finally, we obtain the dynamics reduced to the 
Hamiltonian flow in the 1--dimensional configurational space.
\begin{equation}
\mathcal{H}(a,\dot{a}) = \frac{\dot{a}^{2}}{2} + V(a).
\label{reparam5}
\end{equation}
It can be checked that $\mathcal{H} = E = {\rm const}.$ is a constant of motion provided that energy density $\rho$
satisfies continuity condition (\ref{dyn2}). Therefore, trajectories of the system in the phase space 
$(a,\dot{a})$ lie on the energy level $\mathcal{H} = E = {\rm const}.$ (in the case of the vacuum cosmology we have
$E = 0$). Of course, the Hamiltonian constraint should be consistent with the form of first integral 
(\ref{friedmann1}). Hence, from (\ref{friedmann1}) we obtain
$$\frac{\dot{a}^{2}}{2} + V(a) = \frac{\dot{a}^{2}}{2} + \frac{k}{2} - \frac{\rho a^{2}}{6} - \frac{\Lambda a^{2}}{6},$$
i.e. the potential for the system (\ref{reparam5}) is 
\begin{equation}
2 V(a) = \varphi^{2}(a)\Bigg(k - \frac{\rho}{3}a^{2} - \frac{\Lambda a^{2}}{3}\Bigg) \equiv - \varphi^{2}(a) \frac{\rho_{{\rm eff}}}{3} a^{2}.
\label{reparam6}
\end{equation}
Now, the physical trajectories lie on the zero--energy level $\mathcal{H} = E = 0$ which coincides with the 
form of the first integral, because now the curvature and $\Lambda$--terms are included into the effective 
density $\rho_{{\rm eff}}$. Formally, the curvature term can be absorbed into the potential function by 
postulating the curvature fluid for which $w_{k} = -1/3$, $\rho_{k} = -3k/a^{2}$ (as well as cosmological 
constant term, where $p_{\Lambda} = - \rho_{\Lambda}$, $\rho_{\Lambda} = \Lambda$).

From (\ref{reparam5}) and (\ref{reparam6}) we can see that both Hamiltonians
$$\mathcal{H} = \frac{1}{2}\Bigg(\frac{da}{dt}\Bigg)^{2} + V(q), \hspace{5mm} \bar{\mathcal{H}} = \frac{1}{2}\Bigg(\frac{da}{d \tau}\Bigg)^{2} + \varphi^{2}(a) V(q)$$
reproduce, in the consistent way the same Friedmann equations and they also reproduce equivalent equation of 
motion, because the gauge freedom in choosing the lapse function (i.e. freedom in reparametrization
of time $t \rightarrow \tau$:~$dt = \varphi d\tau$: $\bar{\mathcal{H}} = \varphi^{2}(a) \mathcal{H}$).

Let us consider special case of constant $w$. We have
$$\psi(a) = \frac{1+3w}{2}a^{-1},\hspace{10mm}\varphi = a^{\frac{1+3w}{2}},$$
$$V(a) = \Bigg(\frac{1}{2}k-\frac{\Lambda}{6}a^{2}+\frac{\rho_{0}}{a^{1+3w}}\Bigg)a^{1+3w},$$
$$\mathcal{H} \rightarrow \bar{\mathcal{H}} = a^{1+3w}\Bigg(\frac{1}{2}\dot{a}^{2}+V(a)\Bigg).$$
Classical equations of motion are
\begin{equation}
\dot{a} = \frac{\partial}{\partial \dot{a}}(N \mathcal{H}),\hspace{10mm} \ddot{a} = -\frac{\partial}{\partial a}(N \mathcal{H}),
\label{reparam7}
\end{equation}
where
$$N = a^{1+3w}.$$
The lapse function $N$ plays the role of a Lagrange multiplier and upon its variation we obtain Hamiltonian
constraint $\mathcal{H} = 0$.

The cases of constant quintessential coefficient $w$ are important in applications. The forms of 
potential function for different kinds of matter and flat model are presented in Table \ref{diffmatt}.
\begin{table}[!h]
\caption{The forms of potential function for different kinds of matter and flat model}
\begin{center}
\begin{ruledtabular}
\begin{tabular}{c|c|c|c|c|c|c|c}
 &stiff mat.&rad.&dust&string&top. def.&$\Lambda$&phantom\\
\hline
$w$ &$1$&$1/3$&$0$&$-1/3$&$-2/3$&$-1$&$-4/3$\\
$V$&$\propto a^{-4}$&$\propto a^{-2}$&$\propto a^{-1}$&const.&$\propto a$&$\propto a^{2}$&$\propto a^{3}$\\
\end{tabular}
\end{ruledtabular}
\end{center}
\label{diffmatt}
\end{table}
Of course the presented formalism can be simply generalized to the case of any mixture of noninteracting 
multifluids with the equation of state $p_{i} = w_{i}\rho_{i}$ \cite{szydlowski}. For our purpose it is 
useful to put the dynamical equations into a new form by using dimensionless quantities
\begin{equation}
x \equiv \frac{a}{a_{0}},\hspace{5mm} T \equiv |H_{0}|t,\hspace{5mm}\Omega_{i,0} = \frac{\rho_{i,0}}{\rho_{{\rm cr},0}}
\label{reparam8}
\end{equation}
with
$$H = \frac{\dot{a}}{a},\hspace{5mm}\rho_{{\rm cr},0} = 3H_{0}^{2}.$$
The subscript $0$ means here that the quantity with this subscript is evaluated today (at time $t_{0}$).
Additionaly, we define $\Omega_{k} = -3k/6H_{0}^{2}$ and $\Omega_{\Lambda} = \Lambda/3H_{0}^{2}$.
Then the Hamiltonian is
\begin{equation}
\mathcal{H} = \frac{\dot{x}^{2}}{2} + V(x),
\label{reparam9}
\end{equation}
where
\begin{equation}
V(x) = - \frac{1}{2}\Omega_{k,0} - \frac{1}{2}\sum_{i}\Omega_{i,0}x^{1-3w_{i}}
\label{reparam10}
\end{equation}
which should be considered on the zero--energy level.

The basic dynamical equations are then rewritten as
\begin{gather}
\frac{\dot{x}^{2}}{2} + V(x) = 0,\nonumber \\ 
\ddot{x} = \frac{1}{2}\sum_{i}\Omega_{i,0}(1-3w_{i})x^{-3w_{i}}.
\label{reparam11}
\end{gather}
In the general case, potential function can be obtained from $\rho_{{\rm eff}}$ i.e.
\begin{equation}
V(a) = - \frac{\rho_{{\rm eff}}(a)}{6}a^{2},
\label{reparam12}
\end{equation}
where for the quintessence matter we have
\begin{equation}
\begin{split}
\rho_{{\rm eff}} &= \rho_{\Lambda}+\rho_{k}+\rho_{0}a^{-3}\exp{\Bigg\{-3\int^{a}\frac{w(a)}{a}da\Bigg\}} \\
\rho_{{\rm eff}} &= \Lambda - 3k(1+z)^{2} \\ 
& \quad + \rho_{0}(1+z)^{3}\exp{\Bigg\{3\int^{a}\frac{w(a(z))}{1+z}dz\Bigg\}}.
\end{split}
\label{reparam13}
\end{equation}

\noindent
{\bf 2.} In the second approach to the FRW dynamics representation by the Newtonian equation of motion, we 
redefine the position variable $a$, namely we define X in the following way
\begin{equation}
a \rightarrow X \equiv a^{D(a)},
\label{redef1}
\end{equation}
where $D(a)$ is chosen in such a way that the term with $\dot{a}^{2}$ is absent in (\ref{dyn3}). Hence
we obtain new variable 
\begin{equation}
X = \int^{a}\varphi(a)da,\hspace{5mm} D(a) = \log_{a}{\int^{a}\varphi(a)da} 
\label{redef2}
\end{equation}
and the equation of motion takes the form
\begin{equation}
\ddot{X} = -\kappa(a(X))\varphi(a(X)) = -\frac{\partial V}{\partial X}.
\label{redef3}
\end{equation}
In the special case of constant $w$ we have
$$D(w) = \log_{a}{\int^{a}a^{\frac{1+3w}{2}}da} = \log_{a}{a^{\frac{3}{2}(1+w)}} = \frac{3}{2}(1+w),$$
and because of the obvious formula
\begin{equation}
V(a(X)) = \int^{a}\kappa(a)\varphi(a)dX,
\label{redef4}
\end{equation}
in the considered case, we obtain
\begin{equation}
\varphi = a^{\frac{1+3w}{2}},\hspace{10mm} X = a^{\frac{3}{2}(1+w)}.
\label{redef5}
\end{equation}
$$\mathcal{H}(X,\dot{X}) = \frac{\dot{X}^{2}}{2} + V(X) \equiv 0,$$
\begin{equation}
\begin{split}
V(X) & = \frac{3}{2}(1+w)\Bigg\{\frac{k}{2} - \frac{\Lambda}{6}X^{\frac{2}{D}} + \frac{\rho_{0}}{X^{\frac{1+3w}{D}}}\Bigg\}X^{\frac{1}{D}(1+3w)} \\
& = \frac{3}{2}(1+w) \Bigg\{\frac{k}{2}X^{\frac{2(D-1)}{D}} - \frac{\Lambda}{6}X^{2} + \rho_{0}\Bigg\}.
\end{split}
\label{redef6}
\end{equation}

Let us observe that due to the new variable $X$, the considered system assumes the very simple form.
It can be now considered on the energy level $\mathcal{H} = E > 0$, $E = \frac{3}{2}(1+w)\rho_{0}$ and 
$V(X) = \frac{k}{2}X^{2(1-\frac{1}{D})} - \frac{\Lambda}{6}X^{2}$. This system was analyzed on the phase 
plane $(X,\dot{X})$ many years ago from the point of its structural stability. It is interesting that there 
exists its natural generalization to the case of quintessential cosmology.

Let us now emphasize some advantages of the considered approach. In general, the choice of
\begin{equation}
X \equiv \int^{a}\sqrt{a}\exp{\bigg(\frac{3}{2}\int^{a}\frac{w(a')}{a'}da'\bigg)}da
\label{redef7}
\end{equation}
makes it possible to reduce Q.C. to the nonlinear particle mechanics. There are many advantages of using 
language of nonlinear mechanics. Let us consider some of them.

\noindent
{\bf 1.}
Representation of dynamics as a one-dimensional Hamiltonian flow allows us to make the classification of
possible evolution paths in the phase space as well as in the configuration space. Then the discussion of the
existence and stability of critical points can be performed based on the geometry of the potential function.
It is so because the stability of critical points is determined by the Hessian 
($\partial^{2}\mathcal{H}/\partial x^{i}\partial y^{i}$). In our case the Hamiltonian function assumes the 
very simple form characteristic for simple mechanical systems for which the Lagrangian function is natural, 
i.e. quadratic in velocities $\mathcal{L} = \frac{1}{2}g_{\alpha\beta}\dot{q}^{\alpha}\dot{q}^{\beta}-V(q)$.
In our case, $g_{\alpha\beta} = {\rm const}.$ Then the characteristic equation for linearization of the 
system is $\lambda^{2} + \rm{det}A = 0$, where $\lambda_{i}$ are eigenvalues of the linearization matrix $A$ of 
the Hamiltonian system. Therefore, the only possible critical points in a finite domain of the phase space 
are centres (${\rm det}A > 0$) or saddles (${\rm det}A < 0$). Of course, we can explore dynamics given by the
canonical equations. Denoting $x = a$, $y = \dot{a} = \dot{x}$ we obtain 2D dynamical system in the form
\begin{gather}
\dot{x} = \frac{\partial \mathcal{H}}{\partial y} = y, \nonumber \\
\dot{y} = - \frac{\partial \mathcal{H}}{\partial x} = - \frac{\partial V}{\partial x}.
\label{2ddyn}
\end{gather}

We can observe that trajectories are integrable in quadratures. Namely, from the Hamiltonian constraint 
$\mathcal{H} = E = 0$ we obtain the integral
\begin{equation}
t - t_{0} = \int_{a_{0}}^{a}\frac{da}{\sqrt{-2V(a)}},
\label{2dint}
\end{equation}
and for some specific forms of the potential function we can obtain the exact solutions.

\noindent
{\bf 2.}
It is possible to make the classification  of qualitative evolution paths by analyzing the characteristic 
curve which represents the boundary equation in the configuration space. For this purpose we consider the 
equation of the zero velocity, $\dot{a} = 0$, which represents the boundary of the domain admissible for 
motion $\mathcal{D}_{E} = \{a\in \mathbb{R}_{+}: 2(E-V) \geqslant 0 \}$. By considering the boundary of 
$\mathcal{D}_{E}$ given by the condition 
$$\partial \mathcal{D}_{E} = \{a: V(a) = 0\},$$
full qualitative classification of evolutional paths can be performed.

\noindent
{\bf 3.}
We can find the domains of cosmic acceleration as well as the domains for which the horizon problem is 
solved. Because $\ddot{a} = -dV/da$, one can easily see that acceleration of the universe takes place if 
$V(a)$ is a decreasing function of its argument. The condition $\partial V/\partial a|_{a=a_{0}}$ determines 
the static critical point on the phase plane.

Another interesting question concerns the horizon problem. It is easy to prove the following criterion
of avoiding this problem. The FRW cosmological model does not have an event horizon near the singularity 
if $\dot{a}(t)c^{-1}$ tends to a constant while $a(t)$ tends to zero \cite{weinberg}.\\
 
When all events whose coordinates in the past are located beyond some distance $d_{H}$ then can never 
communicate with the observer at the coordinate $r = 0$ (in R-W metric). We can define the distance
$d_{H}$ as the past event horizon distance
$$d_{H}(t) = a(t) \int_{t_{0}}^{t}\frac{dt'}{a(t')}c = a(t)I.$$
Of course, whenever $I$ diverges as $t_{0} \rightarrow 0$, there is no past event horizon in the spacetime 
geometry \cite{weinberg}. Then it is in principle possible to receive signals from sufficiently early
universe from any comoving particle like a typical galaxy. It the $t'$ integral converges for 
$t_{0} \rightarrow 0$ then our communication with observer at $r=0$ is limited by what Rindler has called a 
particle horizon.
The particle horizon will be present if energy density is growing faster then $a^{-2-\epsilon}$ as 
$a \rightarrow 0$ ($\epsilon \geq 0$)\cite{weinberg}. Therefore if $\dot{a}>a^{-\epsilon/2}$ ($\dot{a}>A$) 
then there is a particle horizon in the past. 

Let $c^{-1}\dot{a} < A$ ($c={\rm const}.$ is assumed but the corresponding theorem can be established for 
the case of variable c(t) \cite{szydlowski}).
Then $I \geqslant \frac{1}{A}\int_{0}^{a_{0}}\frac{da}{a} = \frac{1}{A}(\ln{a_{0}}+\infty)$. Therefore,
$I$ diverges as $a \rightarrow 0$ and there exists no past horizon if the velocity of expansion factor 
$\dot{a} \leqslant A$ is bounded. Our investigation of the particle horizon is independent
of any specific assumption about the behaviour of $a(t)$ near the singularity or of specific form of the 
equation of state. If one assumes a linear equation of state $p = w \rho$ and $w = {\rm const}.$, then 
Friedmann's equations imply the following behaviour for $a(t) \backsimeq (t)^{\frac{2}{3(\gamma+1)}}$ near 
the singularity $t = 0$. The~integral $I$ would thus diverge only if $\gamma < -1/3$, i.e. only if the 
pressure $p$ becomes negative. This is the condition for solving the horizon problem and it is identical to 
that for the solution of the flatness problem ($8 \pi G \rho/3$ term will dominate the curvature term in a 
long time evolution). We can also show here that the integral $\int \frac{dt}{a(t)}$ would diverge only if 
the pressure of the cosmological fluid takes negative values in the general case of $w(a(t))$. 
Conservation condition can be rewritten in the form
$$\frac{dp}{dt} = \frac{d}{dt}[\rho a^{3}(1+w(a))].$$
We can werify that boundedness of $\dot{a}(t)$ means also that $\rho a^{2}$ remains bounded near the
singularity (see FRW eq.). Therefore, $\rho a^{3} \rightarrow 0$ as $t \rightarrow 0$. By integrating both
sides of the equation written above from $0$ to $t$ we obtain
$$-3\int_{0}^{t}pa^{2}\dot{a}dt = a^{3}\rho(t) \geqslant 0.$$
Consequently, $w(a)$ must assume negative value without any specific assumption about the equation of 
state.

The above criterion can be now formulated in the language of the phase space variables. 
If $V(a) \leqslant {\rm const}.$ (it does not diverge at singularity) as $a \rightarrow 0$ then there exists 
no past horizon of the particle.

\noindent
{\bf 4.}
Another natural generalization of the presented formalism consists in including anisotropy in simple Bianchi 
models (Bianchi I or Bianchi V models) or VSL models with variable velocity of light of the 
Albrecht--Maguejo--Barrow type.

In the case of model BI ($k = 0$) or BV ($k = -1$), basic equation assumes the generalized form
\begin{equation}
\frac{\ddot{a}}{a} = -\frac{2}{3}\sigma^{2} - \frac{1}{6}(1 + 3w(a))\rho + \frac{\Lambda c^{2}}{3},
\label{anis1}
\end{equation}
where $c(a) = c_{0}a^{n}$ (in the AMB parametrization).

Equation (\ref{anis1}) has generalized first integral
\begin{equation}
\Bigg(\frac{\dot{a}}{a}\Bigg)^{2} = \frac{\rho}{3} - \frac{kc^{2}}{a^{2}} + \frac{\sigma^{2}}{3} + \frac{\Lambda c^{2}}{3},
\label{anisint}
\end{equation}
where $\sigma^{2} = \frac{1}{2}\sigma_{ab}\sigma^{ab}$ is the shear scalar which can be used to reduce 
dynamical equations to form (\ref{dyn3}), but now $\psi(a)$ preserves its form whereas 
$\kappa(a)~\rightarrow~\bar{\kappa}(a)$ takes the new one which builds the new potential function.
\begin{equation}
\begin{split}
\bar{\kappa}(a) &= \frac{c^{2}(a)}{2}\Bigg[(1+3w(a))\frac{k}{a} - (1+w(a))\Lambda a \Bigg] \\ 
& \quad + \frac{\sigma^{2} a}{2}(1-w(a)), \\
V(a) & \rightarrow  \bar{V}(a) =  \int^{a}\bar{\kappa}(a)\varphi^{2}(a)da.
\end{split}
\label{anis2}
\end{equation}

\noindent
{\bf 5.}
Let us now consider a class of FRW cosmologies with dissipation in the form of bulk viscosity. Owing to
the assumed spacetime FRW symmetries the shear viscosity vanishes, and we deal only with bulk viscosity 
dissipation. It was given by Weinberg who classified the physical significance of bulk viscosity effects 
within the framework of general relativity.

The presence of bulk viscosity is equivalent to introducing effective pressure
\begin{equation}
p_{{\rm eff}} = w(a(z))\rho - 3 \xi H,
\label{viscos1}
\end{equation}
where $\xi$ is the bulk viscosity coefficient (constant for the sake of simplicity).

If we introduce new variable $X$ following previously described scheme, we obtain the generalization of 
(\ref{redef3}) in the form
\begin{equation}
\ddot{X} - \alpha \dot{X} + \frac{\partial V}{\partial X} = 0.
\label{viscos2}
\end{equation}
Equation (\ref{viscos2}) has the form of nonlinear oscillator equation with the damping force 
$\propto \dot{X}$ and the exciting force $F(X) = -\frac{\partial V}{\partial X}$. In the following, this 
system will be analyzed qualitatively by using methods of dynamical system. Therefore, on the basis of 
the corresponding reduction, the particle--like description may be given if effects of bulk viscosity are 
included.

\noindent
{\bf 6.}
If we specialize the standard situation in cosmology with $\Lambda = 0$ and fluid which satisfy weak energy
condition $p + \rho \geqslant 0$ in which we set $p = w\rho$ and $w = {\rm const}.$ then the conservation 
equation gives $\rho \propto a^{-3(1+w)}$ and the $\rho/3$ term will dominate the curvature therm $ka^{-2}$ 
at large $a$ so long as the matter stress violate strong energy condition $\rho + 3p < 0$ and obey 
$p + \rho \geqslant 0$, that is if $-1 \leqslant w \leqslant -1/3$. This is what we shall mean by the 
flatness problem. The scale factor then evolves as $a(t) \propto t^{\frac{2}{3(w+1)}}$ if $w > -1$ or 
$\exp{(H_{0}t)}$, $H_{0}$ constant if $w = -1$.

In the general case of variable $w(a(z))$ quintessential model will provide a solution of flatness problem
if zero curvature solution is representing late time attractor in the phase plane. This is possible in terms
of potential function if $\dot{a} > {\rm const}.$ as $a >> 1$ or
\begin{equation}
-C^{2}/2 < V < 0
\label{solvflat}
\end{equation}
This is the condition oposite to that for the solution of horizon problem $V < -C^{2}/2$ but it is formulate
rather for large $a$ than for small $a$.
Let us notice that if $\ddot{a} > 0$ as $a >> 1$ then $\dot{a} > Ct$ and period of accelerated expansion in 
long term behaviour can solve the flatness problem.

Let us consider the case of non-zero cosmological constant added to the potential function. Then in order
to explain why it does not totally dominate matter term $\rho/3$ at large $a$ we would need the presence of 
fluid with subnegative pressure, i.e. with $w < -1$ such that a weak energy condition is violated 
$p + \rho < 0$. This is what we shall mean by the cosmological constant problem. If we assume that $\rho$
is decreasing function of scale factor the solution of cosmological constant problem leads to contradiction
$\dot{a} < 0$. 
In the general case we have following condition for solving $\Lambda$-problem
\begin{equation}
\rho_{{\rm eff}} > \Lambda \Leftrightarrow \frac{d}{da}\Big(\frac{V}{a^{2}}\Big) > 0 \Leftrightarrow 
|V| > \Lambda a^{-2}
\label{solvlambda}
\end{equation}
i.e. modulus of potential function is growing faster than $a^{-2}$ at large $a$.

Let us notice that for $w = {\rm const}.$ above condition leads to the identical condition to that to solve 
the horizon problem $-1 \leqslant w \leqslant -1/3$. It is consequence of the fact that corresponding 
conditions are valid at different domains $a \rightarrow \infty$ and $a \rightarrow 0$.

\section{Applications}

We assume now that quintessential coefficient $w(a(z)) \equiv p/\rho$ is the analytical function 
of~$z$
\begin{equation}
w(a(z)) = w(0) + \frac{dw}{dz}\Bigg|_{0}z + \frac{1}{2}\frac{d^{2}w}{dz^{2}}\Bigg|_{0}z^{2} + \dotso
\label{wrozw}
\end{equation}
Let $\gamma_{0} = w(0)$ and 
$$\gamma_{1} = \frac{dw}{dz}\Bigg|_{0},\hspace{5mm} \gamma_{2} = \frac{1}{2}\frac{d^{2}w}{dz^{2}}\Bigg|_{0},
\hspace{2mm} \dotso, \hspace{2mm} \gamma_{i} = \frac{1}{i!}\frac{d^{i}w}{dz^{i}}\Bigg|_{0}.$$
We consider models with vanishing $\Lambda$--term because conception of quintessence is treated as an 
alternative to the cosmological constant. The potential function of the dynamical system describing the FRW 
model with equation of state (\ref{wrozw}) is given by
\begin{equation}
\begin{split}
V & = \int^{a}\kappa(a)\varphi^{2}(a)da = \frac{k}{2}a\exp{\Bigg\{3 \int^{a}\frac{w(a)}{a}da\Bigg\}} \\
  & = \frac{k}{2}(1+z)^{-1}\exp{\Bigg\{(-3) \int^{z}\frac{w(a(z))}{(1+z)}dz\Bigg\}}.
\end{split}
\label{potqc}
\end{equation}
For example, in the case of mixture of noninteracting dust and fluid for which 
$p_{x} = w_{x}\rho_{x}$, where $w_{x} = {\rm const}.$, we have
\begin{equation}
V(a) = \frac{k}{2}a\Bigg(\frac{a^{3w_{x}}}{\frac{\rho_{0m}}{\rho_{0x}}a^{3w_{x}} + 1}\Bigg),
\label{potqc1}
\end{equation}
here we consider quintessential matter for which
$$\rho_{x} = \rho_{0x} a^{-3(1+w_{x})}, \hspace{5mm} \rho_{0x} = {\rm const}.$$
and dust matter for which
$$\rho_{m} = \rho_{0m} a^{-3}, \hspace{5mm} \rho_{0m} = {\rm const}.$$
If for quintessential fluid $w_{x} = w_{x}(a)$, then the potential function takes the more general form
\begin{equation}
V(a) = \frac{k}{2}a\Bigg[\frac{\exp{\Big(3\int^{a}\frac{w_{x}(a)}{a}da}\Big)} {\frac{\rho_{0m}}{\rho_{0x}}
\exp{\Big(3\int^{a}\frac{w_{x}(a)}{a}da}\Big)  + 1}\Bigg].
\label{potqc2}
\end{equation}

After substituting (\ref{wrozw}) into (\ref{potqc}) we obtain the general form of the potential function 
which can be useful for a classification of cosmological models in the configurational space
\begin{equation}
\begin{split}
V(a(z)) &= \frac{k}{2}(1+z)^{-(1+3\gamma_{0})} \\ 
&\times \exp{\Bigg\{(-3)\int^{z}\frac{(\gamma_{1}z + \gamma_{2}z^{2} + \dotso)}{(1+z)}dz\Bigg\}}.
\end{split}
\label{potqc4}
\end{equation}
Let us consider a few special cases of (\ref{potqc4}).

\noindent
{\bf A.}

Let $\gamma_{0} = -\frac{1}{3}$ (like for a string) which determines the boundary the strong
energy condition violation. Additionaly, we consider expansion of $w(z)$ up to the second order term. Then 
we obtain from the integral
\begin{equation}
\int \frac{\gamma_{1}z + \gamma_{2}z^{2}}{1+z}dz = (\gamma_{1}-\gamma_{2})z + \frac{\gamma_{2}}{2}z^{2} 
- (\gamma_{1}-\gamma_{2})\ln{(1+z)}
\label{potqc5}
\end{equation}
the exact form of the potential function
\begin{equation}
V(a(z)) = \frac{k}{2} e^{-3(\gamma_{1}-\gamma_{2})z - \frac{3}{2}\gamma_{2}z^{2}}(1+z)^{3(\gamma_{1}-\gamma_{2})},
\label{potqc6}
\end{equation}
or for any $\gamma_{0}$
\begin{equation}
V(a(z))=\frac{k}{2}e^{-3(\gamma_{1}-\gamma_{2})z-\frac{3}{2}\gamma_{2}z^{2}}(1+z)^{-(1+3\gamma_{0})+3(\gamma_{1}-\gamma_{2})}.
\label{potqc7}
\end{equation}
\begin{figure}[!ht]
\begin{center}
\includegraphics[scale=0.6, angle=0]{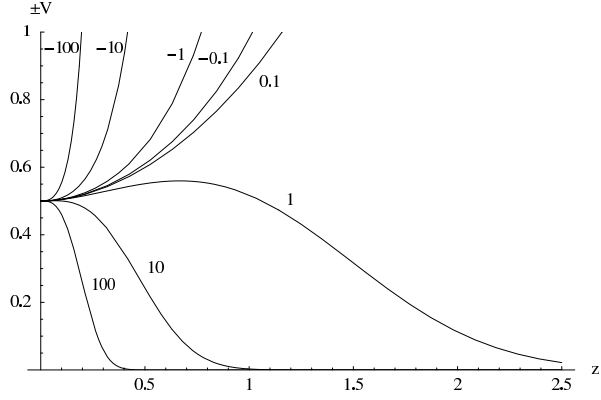}
\caption{Diagram of V(z) (formula (\ref{potqc7})) for the case of $k = \pm 1$, 
$\gamma_{0} = -1/3$, $\gamma_{1} = -2/3$ and various $\gamma_{2}$ 
($\gamma_{2}=-100,-10,-1,-0.1,0.1,1,10,100$)}
\label{wyk1}
\end{center}
\end{figure}

Of course, the trajectories of the considered system lie in the domain admissible for motion defined as
$$\mathcal{D}_{E} = \{a \in \mathbb{R}_{+}: 2(E-V(a)) \geqslant 0, E = \frac{\rho_{0}}{6}\},$$
$$\rho = \rho_{0} a^{-3} \exp{\Bigg(3 \int \frac{w(a)}{a}da \Bigg).}$$
Therefore, equation of the boundary $\partial \mathcal{D}_{E}$ is
$$V=E  \Leftrightarrow \  \frac{k}{2}e^{-3(\gamma_{1}-\gamma_{2})z-\frac{3}{2}\gamma_{2}z^{2}} 
= E (1+z)^{(1+3\gamma_{0})-3(\gamma_{1}-\gamma_{2})}.$$
If we substitute $\gamma_{0} = -\frac{1}{3}$, then we obtain the potential function expressed in terms of 
a single parameter $\gamma_{1}/\gamma_{2} \equiv x$, namely
$$\Bigg(\frac{1+z}{e^{z}}\Bigg)^{(x-1)} = \Bigg(\frac{2E}{k}\Bigg)^{\frac{1}{3\gamma_{2}}}e^{z^{2}/2} 
= \bar{E}e^{z^{2}/2}.$$
Finally, we obtain
\begin{equation}
x(z) = 1 + \frac{\ln{\bar{E}} + z^{2}/2}{\ln{(1+z)}-z},
\label{potqc8}
\end{equation}
where $\bar{E} = (2E/k)^{1/3\gamma_{2}}$.\\
\begin{figure}[!ht]
\begin{center}
\includegraphics[scale=0.6, angle=0]{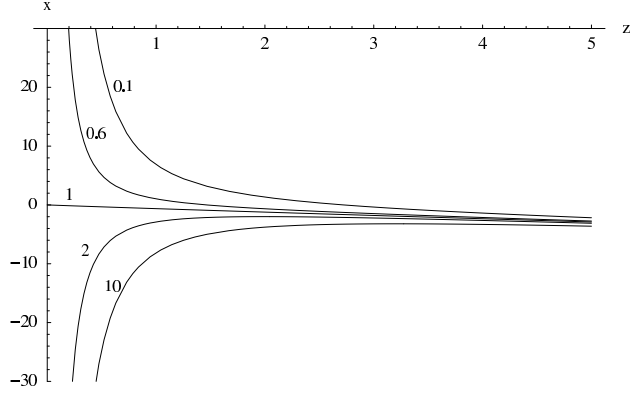}
\caption{Diagram of dependence $x(z)$ for classification of models with potential (\ref{potqc7}) 
($\bar{E}~=~0.1,0.6,1,2,10$)}
\label{wyk2}
\end{center}
\end{figure}
The plot of $x(z)$ for different $\bar{E}$ is shown in Fig. \ref{wyk2}.  

Let us note that if $\rho_{0} > 0$
($E > 0$), then there is no boundary for $k = -1$, whereas if $\rho_{0} < 0$ ($E < 0$), there is
no boundary for $k = +1$ of the domain admissible for motion. Finally, we consider the evolution path
as a level of $x = {\rm const}.$ and then we classify all evolutions modulo the quantitative properties of 
their dynamics \cite{robertson}.

\noindent
{\bf B.}

\begin{figure}[!ht]
\begin{center}
\includegraphics[scale=0.6, angle=0]{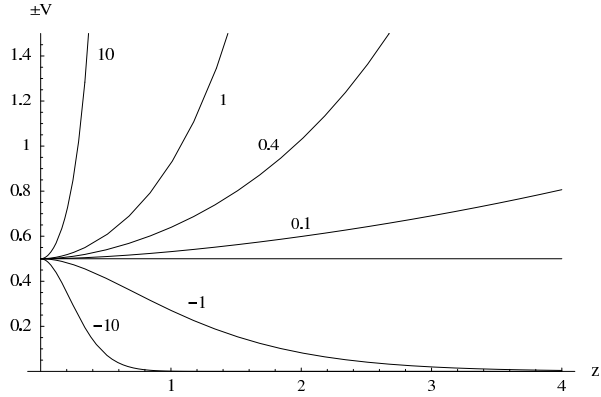}
\caption{Diagram of the potential function $V(z)$ for the case {\bf B},  
$p~=~(\gamma_{0}+\gamma_{1}z)\rho$, $\gamma_{0} = -1/3$, $\gamma_{1} = -2/3\beta$ 
($\beta = -10,-1,0.1,0.4,1,10$), $k = \pm 1$}
\label{wyk3}
\end{center}
\end{figure}

\begin{figure}[!ht]
\begin{center}
\includegraphics[scale=0.6, angle=0]{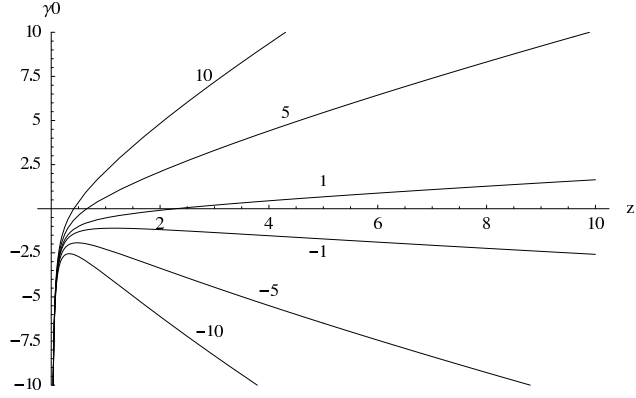}
\caption{Diagram of boundary curve $\partial \mathcal{D}_{E}$ usefull in classification of the model
in the configurational space (formula (\ref{potqc10})). We put here $\gamma_{1} = -2/3\beta$, 
($\beta = -10,-5,-1,1,5,10$), $c = -1$}
\label{wyk4}
\end{center}
\end{figure}

\begin{figure}[!ht]
\begin{center}
\includegraphics[scale=0.6, angle=0]{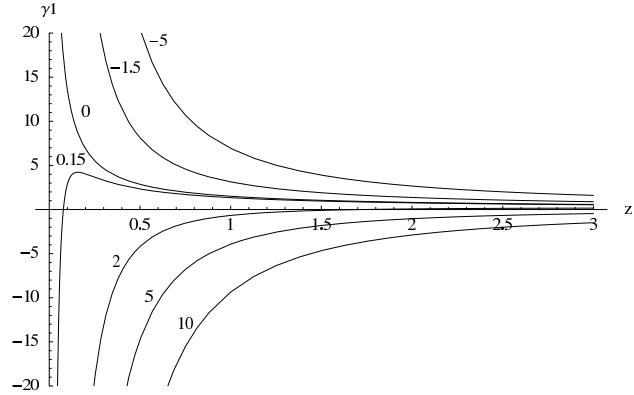}
\caption{The plot of $\gamma_{1}(z(a))$ (formula (\ref{potqc11})) for different signs of c 
($c = -5,-1.5,0,0.15,2,5,10$)}
\label{wyk5}
\end{center}
\end{figure}

$\gamma_{2} = 0$, $\gamma_{0} \neq 0$, i.e. we consider quintessential model with pressure 
$p~=~\gamma_{0}\rho + (\gamma_{1}z)\rho$. Then we obtain the potential in the form (Fig.\ref{wyk3})
\begin{equation}
V(a(z)) = \frac{k}{2}e^{-3\gamma_{1}z}(1+z)^{-(1+3\gamma_{0})+3\gamma_{1}}.
\label{potqc9}
\end{equation}
Let us consider the boundary of the configuration space given by $V(z)$. Then $\gamma_{0}$ can be expressed 
as a function of $z$ in the following way
\begin{equation}
\gamma_{0}(z) = (-\frac{1}{3}+\gamma_{1}) - \frac{\gamma_{1}z}{\ln{(1+z)}} + \frac{c/3}{\ln{(1+z)}},
\label{potqc10}
\end{equation}
where $z = a^{-1}-1$, $c = -\ln{|\frac{2E}{k}|}$.
The plot of $\gamma_{0}(z)$ for different $\gamma_{1}$ is shown in Fig. \ref{wyk4} for $\rho_{0} > 0$
and $k = +1$.
If we put $\gamma_{0} = -1$ (cosmological constant) then we can obtain classification of cosmological models
with cosmological constant in terms of levels of $\gamma_{1} = {\rm const}.$, where
\begin{equation}
\gamma_{1}(z) = \frac{-\frac{2}{3}\ln{(1+z)}-c/3}{\ln{(1+z)}-z}.
\label{potqc11}
\end{equation}
The plot of $\gamma_{1}(z(a))$, for different $c$, is shown in Fig. \ref{wyk5}.

\noindent
{\bf C.}

\begin{figure}[!ht]
\begin{center}
\includegraphics[scale=0.6, angle=0]{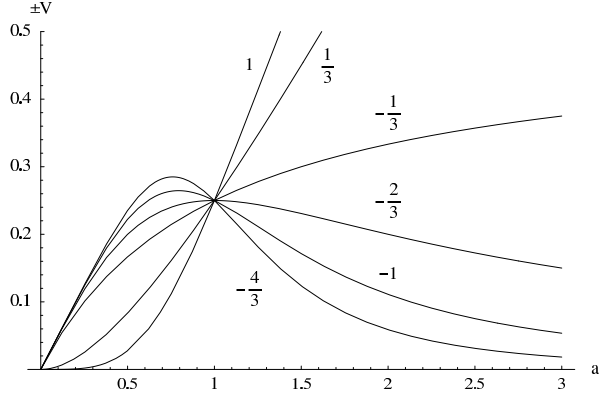}
\caption{Diagram of the potential function $V(a)$ for the case {\bf C}, $p~=~w_{x}\rho_{x}$, for different  
($w_{x}=-\frac{4}{3},-1,-\frac{2}{3},-\frac{1}{3},1,\frac{1}{3}$), $k = \pm 1$}
\label{wyk6a}
\end{center}
\end{figure}

\begin{figure}[!ht]
\begin{center}
\includegraphics[scale=0.6, angle=0]{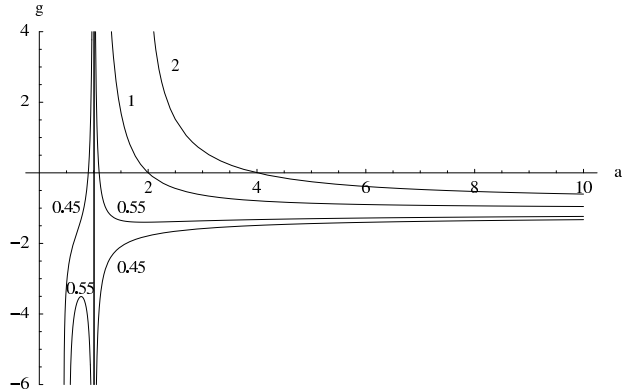}
\caption{Diagram of $g(a)$ for classification of evolution in the configurational space for the case
of constant $w_{x}$ ($A~=~0.45,0.55,1,2$), $\frac{\rho_{0m}}{\rho_{0x}}=1$}  
\label{wyk6}
\end{center}
\end{figure}

$p = w_{x}\rho_{x}$, $w_{x} = {\rm const}.$ 
Let us consider quintessential matter in the form of noninteracting mixture of dust and matter described 
by the equation of state $p = w_{x}\rho_{x}$, where $w_{x} = {\rm const}.$ In this case, classification of 
all evolutional paths can be given in terms of potential function (Fig. \ref{wyk6a})
\begin{equation}
V(a) = \frac{k}{2}a\Bigg(\frac{a^{3w_{x}}}{\frac{\rho_{0m}}{\rho_{0x}}a^{3w_{x}} + 1}\Bigg),
\label{potqc12}
\end{equation}
where
$$\rho_{x} = \rho_{0x} a^{-3(1+w_{x})}, \hspace{5mm} \rho_{m} = \rho_{0m} a^{-3}.$$
The boundary curve $V(a) = E$ can be used to classify all evolutional paths in the configurational
space, namely
\begin{equation}
g(a) = 3w_{x}(a) = \frac{\ln{\bigg(\frac{A}{a - A \frac{\rho_{0m}}{\rho_{0x}}}\bigg)}}{\ln{a}},
\label{potqc13}
\end{equation}
where $A = 2E/k$.\\
The plot of $g(a)$, for different $A$, is shown in Fig. \ref{wyk6}. \\

\noindent
{\bf D.}

Since the observations of the supernovae of type Ia indicate that the universe must be today in an 
accelerated expansion the nature of the fluid responsible for such a behaviour has been object of many 
studies. While the most obvious candidate for such component is the vacuum energy the posibility that the 
dark energy might be described by the Chaplygin gas is seriously suggested \cite{kambenalc}. 

The Chaplygin gas has an interesting motivation connected with the string theory. If we consider a d--brane
configuration in the d+2 Nambu-Goto action, the employment of the light-cone parametrization leads to the
action of a Newtonian fluid with the equation of state $p = -A/\rho$, whose symmetries are the same
as those of the Poincare' group. Hence, the relativistic character of the action is somehow hidden in the 
equation of state (for review see R. Jackiv, A particle field theorist's lectures on supersymmetric, 
non-abelian fluid mechanics and d-branes physics).

Let us consider FRW model with fluid in the form of generalized Chaplygin gas for which equation of state 
is given by
\begin{equation}
p = -\frac{A}{\rho^{\alpha}},\hspace{5mm} 0 \leqslant \alpha < 1.
\label{chapeqst}
\end{equation}
The energy-momentum conservation implies that Chaplygin gas density depends on the scale factor as
\begin{equation}
\rho = \bigg(A + \frac{B}{a^{3(1+\alpha)}}\bigg)^{\frac{1}{1+\alpha}}, 
\label{chapdens}
\end{equation}
where $A$ and $B$ are constants and the Chaplygin gas corresponds to the case $\alpha = 1$; $A$ is a positive
constant because sound velocity of Chaplygin gas is $v_{s}^{2}/c^{2} = A^{2}/\rho^{2}$.

Equation (\ref{chapdens}) interpolates smoothly between a dust dominated phase, where
$\rho \propto a^{-3}$, and a De Sitter phase, where $p = -\rho$, through an intermediate regime described by 
the Zeldovich stiff matter $p = \rho$.

\begin{figure}[!ht]
\begin{center}
\includegraphics[scale=0.6, angle=0]{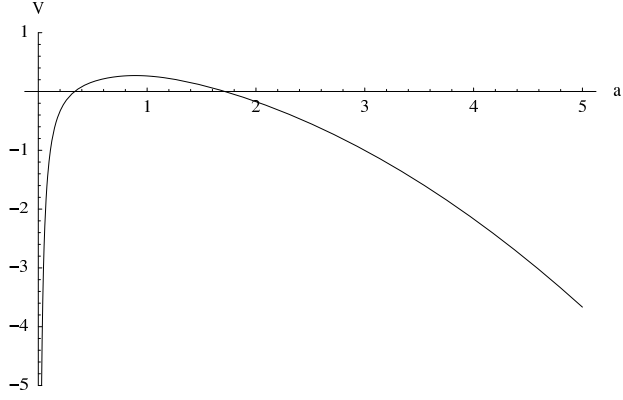}
\caption{Diagram of the potential function $V(a)$ for the case {\bf D} (formula (\ref{chapham})),
$k = A = B = \alpha = 1$.}
\label{wyk7}
\end{center}
\end{figure}

\begin{figure}[!ht]
\begin{center}
\includegraphics[scale=0.6, angle=0]{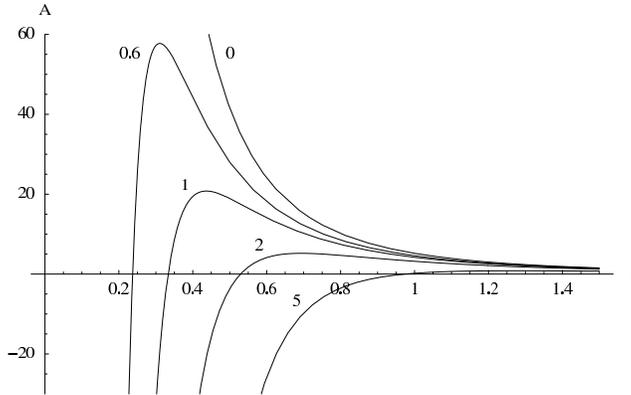}
\caption{Diagram of $A(a)$ for classification of possible evolution paths for the case
of constant $k=+1$, $\alpha = 0.5$ and different $B$ ($B~=~0,0.6,1,2,5$)}
\label{wyk8}
\end{center}
\end{figure}

\begin{figure}[!ht]
\begin{center}
\includegraphics[scale=0.6, angle=0]{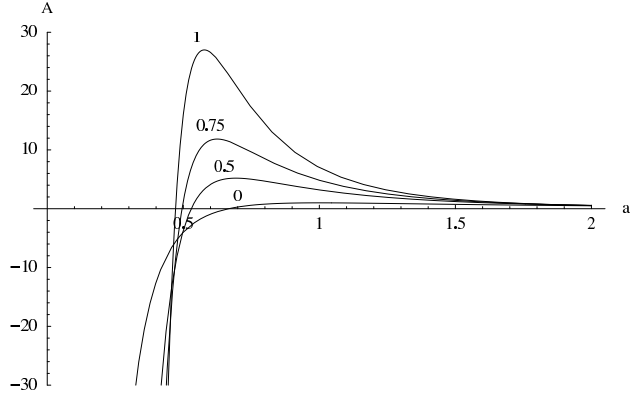}
\caption{Diagram of $A(a)$ for classification of possible evolution paths for the case
of constant $k=+1$, $B = 2$ and different $\alpha$ ($\alpha~=~0,0.5,0.75,1$)}
\label{wyk9}
\end{center}
\end{figure}

Following our previous consideration, the dynamics is given by the hamiltonian
$$\mathcal{H} = \frac{\dot{a}^{2}}{2} + V(a) \equiv 0,$$
where
\begin{equation}
V(a) = -\frac{1}{6}\rho_{{\rm eff}}a^{2} = 
-\frac{1}{6}\bigg(A + \frac{B}{a^{3(1+\alpha)}}\bigg)^{\frac{1}{1+\alpha}}a^{2} + \frac{k}{2}.
\label{chapham}
\end{equation}
The dependence of $V(a)$ for $\alpha = 1$ is illustrated on the Fig. \ref{wyk7}
The domain admissible for motion of the system with generalized Chaplygin gas is
$$\mathcal{D}_{0} = \{a: V(a) \leqslant 0\}.$$
The boundary curve $\partial \mathcal{D}_{0}$ can be used to classify possible evolution paths in 
the configurational space in the following way. We consider constant levels of $A(a)$ relation given by
\begin{equation}
A(a) = \bigg(\frac{3k}{a^{2}}\bigg)^{1+\alpha} - \frac{B}{a^{3(1+\alpha)}}.
\label{chapa}
\end{equation}
The zero velocity curve for $k = +1$ are shown on Fig. \ref{wyk8} and Fig. \ref{wyk9}.
\begin{figure}[!ht]
\begin{center}
\includegraphics[scale=0.35, angle=270]{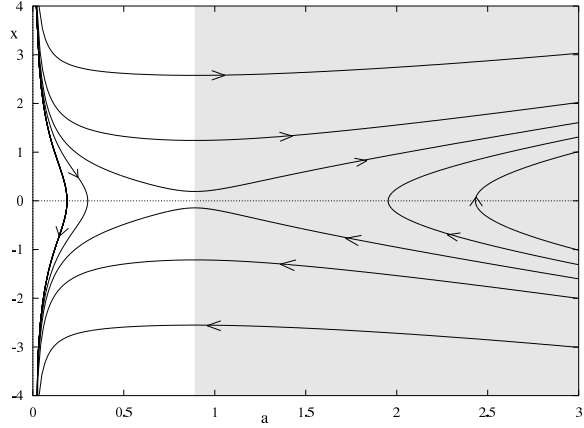}
\caption{The phase portrait $(a,x)$ for case {\bf D} ($k = A = B = \alpha = 1$). The shaded region is a 
region of accelerated expansion of the universe.}
\label{wyk10}
\end{center}
\end{figure}

It is interesting that the existence and character of the critical points of the considered system
$\dot{a} = x,\hspace{3mm} \dot{x} = -\frac{\partial V}{\partial a}$ depends on the geometry of the potential
function.

It can be easily to shown that in our case on a finite region of the phase plane $(a, x)$ only saddle points
are admitted because $\partial^{2} V/\partial a^{2} < 0$ at the critical point $a = a_{0}$, 
$\partial V/\partial a|_{a=a_{0}} = 0$. Therefore, all points are hiperbolic ($TrA = 0$) and the system is 
structurally stable.

In terms of $V(a)$ the domains of accelerating trajectories can be easily found, namelly if 
$\partial V/\partial a < 0$ then the system starts to accelerate. Because the diagram of $V(a)$ is upper 
convex, the static critical point will separate regions without acceleration from the domain in the phase 
space where trajectories accelerate. One can shown that the critical value of $a$ is
$$a = a_{crit} = \Bigg(\frac{B}{2A}\Bigg)^{\frac{1}{3(1+\alpha)}}.$$
At this point the diagram $V(a)$ has the maximum.

\noindent
{\bf E.}

Interesting formulas for $w(a)$ were already proposed by Corasaniti and Copeland \cite{corasaniti}.
They considered a broad class of tracking potentials for scalar fields, namely $V(\phi) \propto 
\phi^{-\alpha}$ -- inverse power low potential (INV) \cite{zlatev}, 
$V(\phi) \propto \phi^{-\alpha} \exp{(\phi^{2}/2)}$ -- supergravity potential (SUGRA) \cite{brax},
$V(\phi) \propto \exp{(-\alpha \phi)} + \exp{(\beta \phi)}$ (2EXP) \cite{barreiro}, The Skordis model 
\cite{albrecht} and Copeland-Nunes-Rosati model (CNR) \cite{copeland}.

\begin{figure}[!ht]
\begin{center}
\includegraphics[scale=0.4, angle=270]{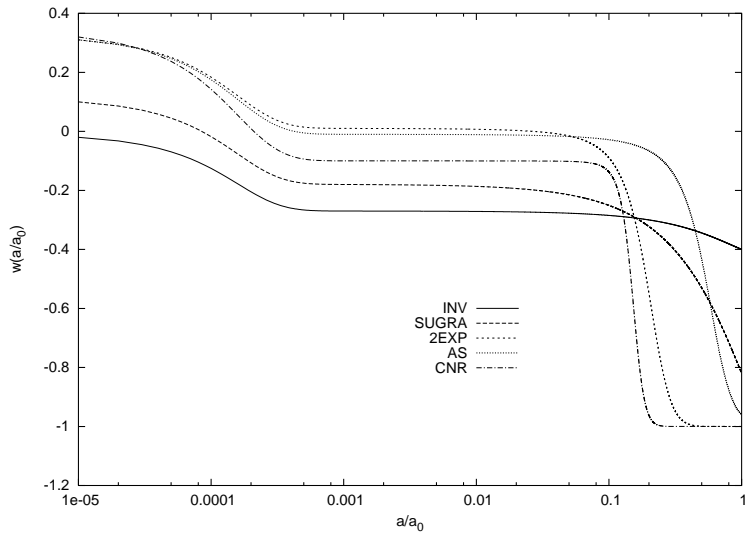}
\caption{The dependence of the equation of state factor $w(a)$ versus the scale factor $a/a_{0}$ for 
different models of potential of scalar field.}
\label{wfig}
\end{center}
\end{figure}

\begin{figure}[!ht]
\begin{center}
\includegraphics[scale=0.4, angle=270]{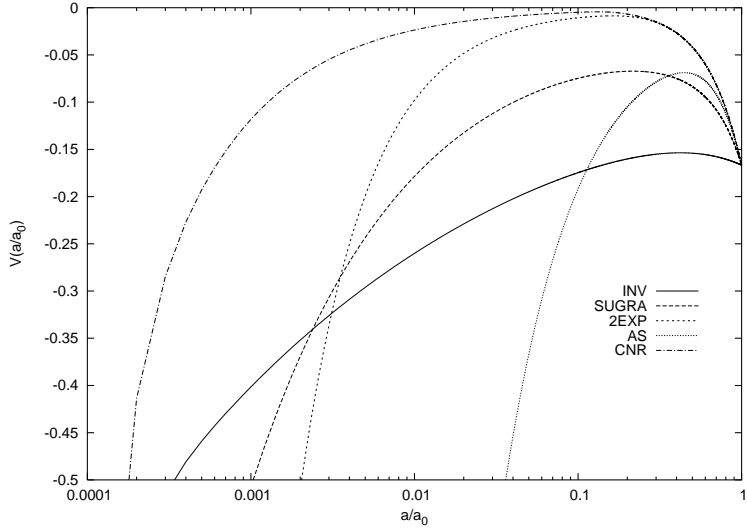}
\caption{The dependence of the potential of the Hamiltonian system against the scale factor $a/a_{0}$ for 
a different class of tracking potentials.}
\label{vfig}
\end{center}
\end{figure}

First, we aply our method to construct a potential function of the corresponding Hamiltonian dynamical 
system. It assumes the following form
\begin{align}
V(a)&=-\frac{\rho_{0}}{6}a^{-1}\exp{\Bigg(3\int_{a}^{1}\frac{w(a)}{a}da\Bigg)}\nonumber \\
&=-\frac{\rho_{0}}{6a}\exp{\Bigg\{3\Bigg(F_{1}\int_{a}^{1}\frac{f_{r}(a)}{a}da+F_{2}\int_{a}^{1}\frac{f_{m}(a)}{a}da+F_{3}(1-a)\Bigg)\Bigg\}},
\label{revpot}
\end{align}
where
\begin{equation}
\int_{a}^{1}\frac{f_{r,m}}{a}da = \sum_{n=0}^{\infty}[{\rm Ei}(-\beta n x)_{r,m}-{\rm Ei}(1)],
\label{eifunc}
\end{equation}
where $x=a-a_{c}$, $\beta=1/\Delta$, Ei -- exponent integral function
and coefficients $F_{1}$, $F_{2}$, $F_{3}$ are determined by the condition that $w(a)$ takes on the 
respective values of $w^{r}$, $w^{m}$ and $w^{0}$, during radiation epoch ($a=a_{r}$), matter domination 
($a=a_{m}$) as well as today ($a=a_{0}=1$).

The coresponding forms of $w(a)$ function for a different class of potentials are illustrated in 
Fig.~\ref{wfig}.

The function $f_{r,m}(a)$ has the following form \cite{corasaniti}
\begin{equation}
f_{r,m}(a)=\frac{1}{1+\exp{[-(a-a_{c}^{r,m})/\Delta_{r,m}]}},
\label{frmform}
\end{equation}
where the corresponding values of coefficients $a_{c}^{r,m}$ are taken from Table I in ref \cite{corasaniti}.

After substitution formulas (\ref{frmform}) to (\ref{revpot}) we obtain different forms of 
potentials for a different class of models (see Fig.~\ref{vfig}). From this figure we can observe that in 
all cases a potential function is of the same type (upper convex), like for the Chaplygin gas. Therefore 
the phase portraits determined from the potential functions of the systems are topologically equivalent. 
From the physical point of view it means that considered models can be seriously treated as candidates for dark energy 
description.

Let us notice that the dynamical system
$$\dot{a} = x, \hspace{5mm} \dot{x} = -\frac{\partial V}{\partial a}$$
can be transformed to $(z,x)$ variables ($a = (1+z)^{-1}$) and then we obtain
$$\frac{dz}{d\tau} = -(1+z)^{2}x,\hspace{5mm} \frac{dx}{d\tau} = (1+z)^{2}\frac{\partial V}{\partial z}.$$
Therefore after the reparametrization of time along trajectories $\tau \rightarrow \eta: d\eta = -(1+z)^{2}d\tau$
(now $\eta$ will be a decreasing function of time variable $\tau$) we obtain the dynamical system
\begin{equation}
\frac{dz}{d\eta} = x, \hspace{5mm} \frac{dx}{d\eta} = -\frac{\partial V}{\partial z}.
\label{qcdynzxeta}
\end{equation}
Of course, system (\ref{qcdynzxeta}) can be analysed in terms of dynamical systems, i.e. in terms of the 
method of qualitative analysis of differential equations on the phase plane $(z,x)$. System 
(\ref{qcdynzxeta}) has the first integral in the form
\begin{equation}
\Bigg(\frac{dz}{d\eta}\Bigg)^{2} - 2(E - V(z)) = 0.
\label{qcdynzxetaint}
\end{equation}
For the considered system only two types of critical points can appear, namelly centres or saddle points.
If ${\rm det}A = \frac{\partial^{2} V}{\partial z^{2}}\big|_{z=z_{0}}$ is negative, the diagram of the 
potential function $V(a(z))$ has maxima; they correspond to the saddle point. On the other hand, if the 
diagram of the potential function $V(a(z))$ has minima, they correspond to centres. It is important that we  
can discuss the stability of critical points based only on the geometry of the potential function. It is easy
to check that at the critical points of the system appearing at $a_{0}~=~{\rm const}. 
(z_{0} = {\rm const}.)$, $\frac{\partial V}{\partial a}\big|_{a_{0}} = 0$ 
($\frac{\partial V}{\partial z}\big|_{z_{0}} = 0$) we have $\kappa(a) = 0$, and then
\begin{equation}
\begin{split}
\frac{\partial^{2} V}{\partial a^{2}}\Bigg|_{a=a_{0}} 
&= \varphi^{2}\frac{3}{2}\frac{k}{a}\frac{dw}{da}\Bigg|_{a=a_{0}} 
= \frac{3}{2}\frac{\varphi^{2}}{a}k\frac{dw}{da}\Bigg|_{w(a)=-1/3} \\
&= -\frac{3}{2}\varphi^{2}k(1+z)^{3}\frac{dw}{dz}\Bigg|_{z=z_{0}}.
\end{split}
\label{qcdynzxeta1}
\end{equation}
Therefore, if $k \frac{dw}{dz}(z_{0})$ is positive, only saddles points can appear which guarantee the 
structural stability of the system.

\begin{figure}[!hp]
\begin{center}
\includegraphics[scale=0.3, angle=270]{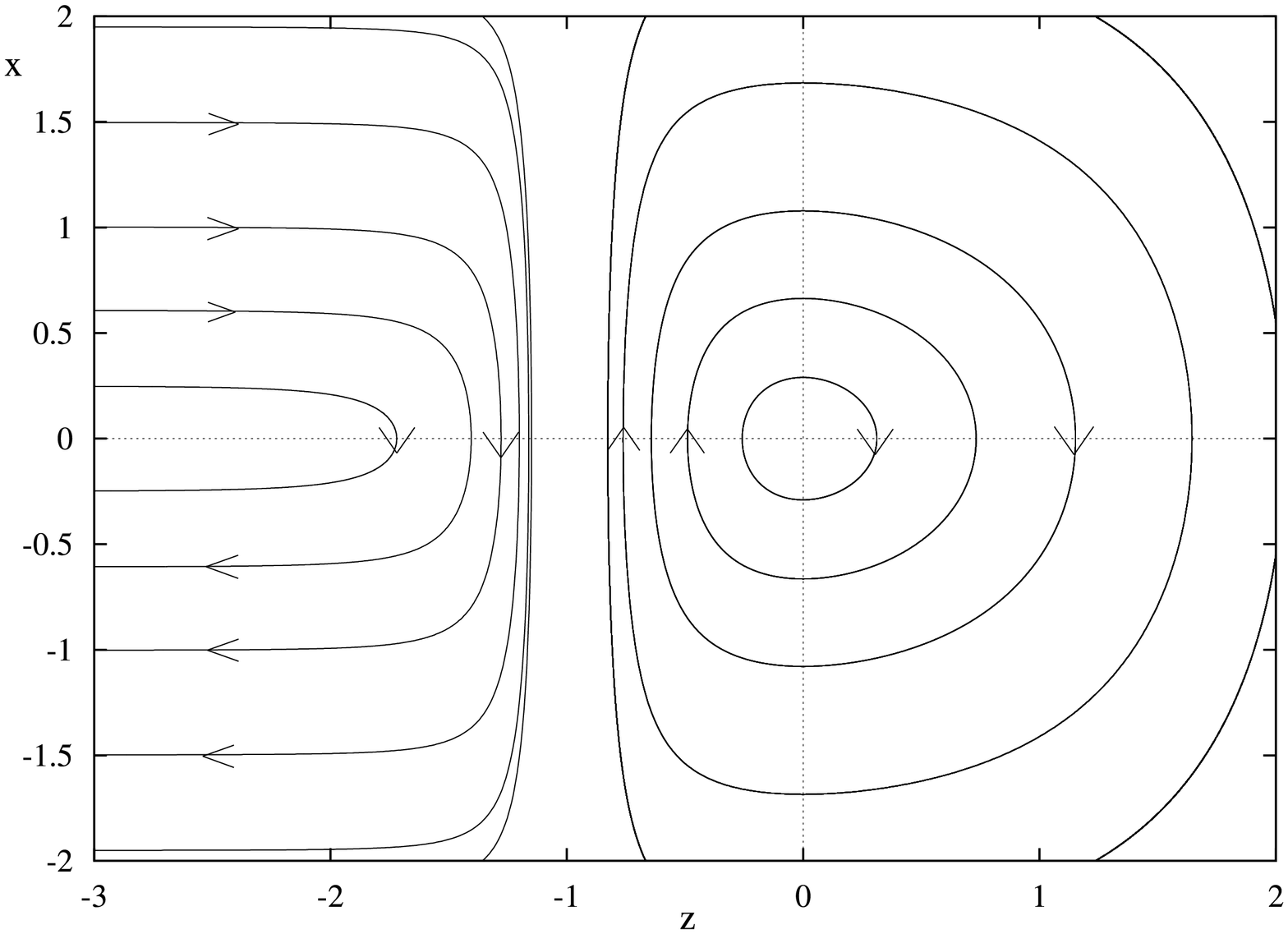}
\caption{The phase portrait $(z,x)$ for case {\bf B} $p = (\gamma_{0} + \gamma_{1}z)\rho$ 
($k = 1$, $\gamma_{0}=-1/3$, $\gamma_{1}=-2/3$)}
\label{B}
\end{center}
\end{figure}

\begin{figure}[!hp]
\begin{center}
\includegraphics[scale=0.3, angle=270]{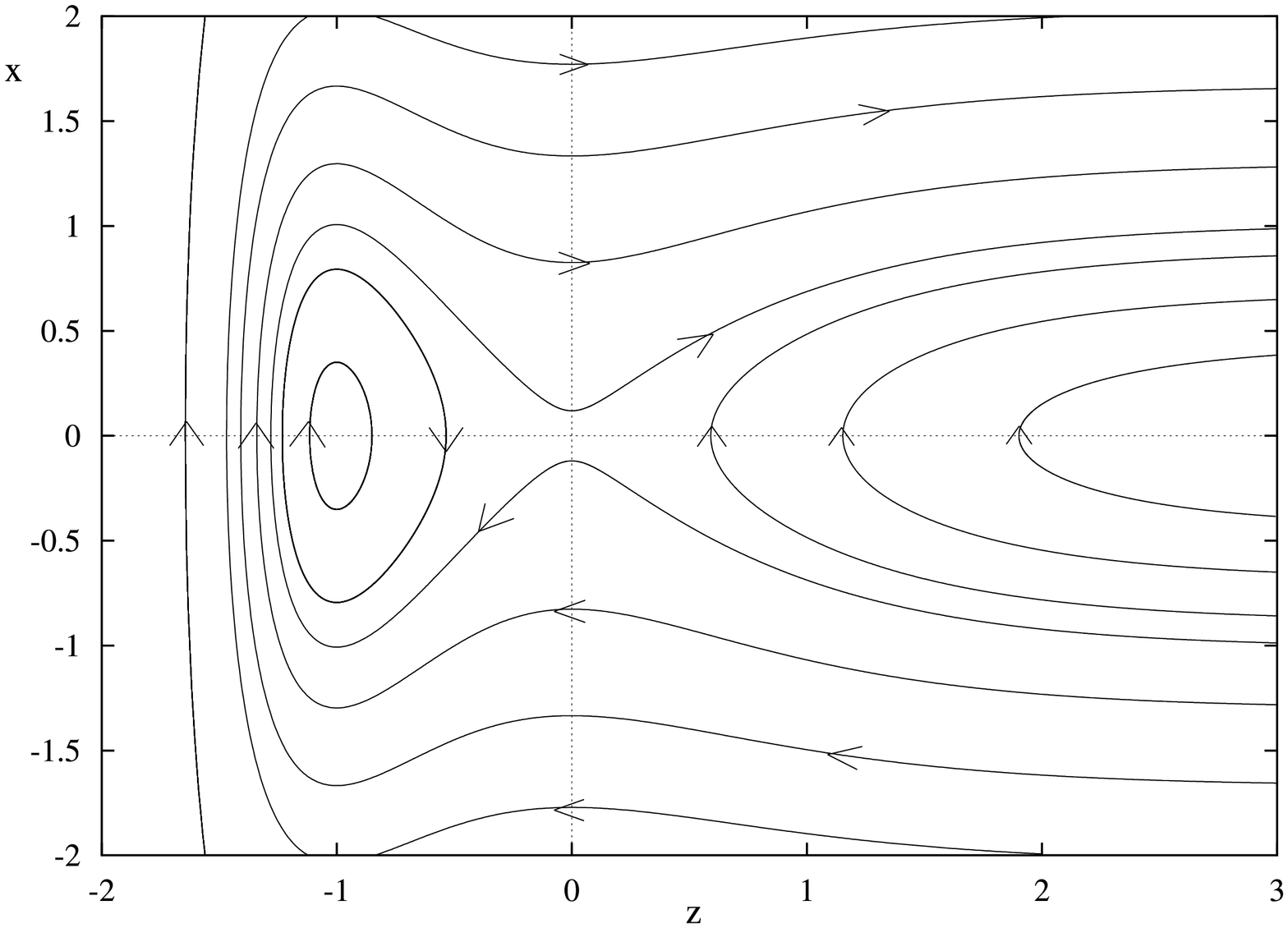}
\caption{The phase portrait $(z,x)$ for case {\bf B} $p = (\gamma_{0} + \gamma_{1}z)\rho$ 
($k = 1$, $\gamma_{0}=-1/3$, $\gamma_{1}=2/3$)}
\label{B1}
\end{center}
\end{figure}

\begin{figure}[!hp]
\begin{center}
\includegraphics[scale=0.3, angle=270]{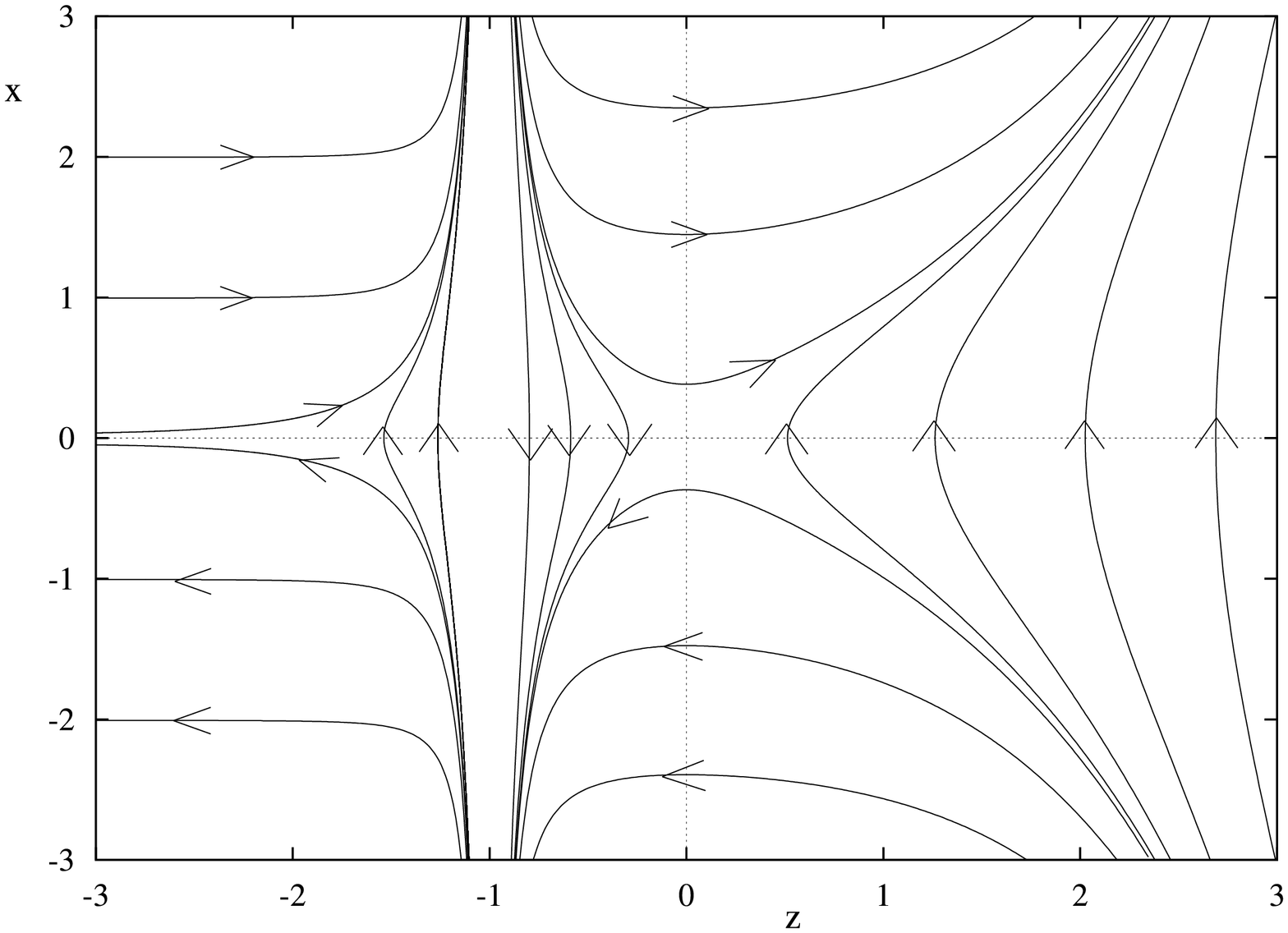}
\caption{The phase portrait $(z,x)$ for case {\bf B} $p = (\gamma_{0} + \gamma_{1}z)\rho$
($k = -1$, $\gamma_{0}=-1/3$, $\gamma_{1}=-2/3$)}
\label{B2}
\end{center}
\end{figure}

\begin{figure}[!hp]
\begin{center}
\includegraphics[scale=0.3, angle=270]{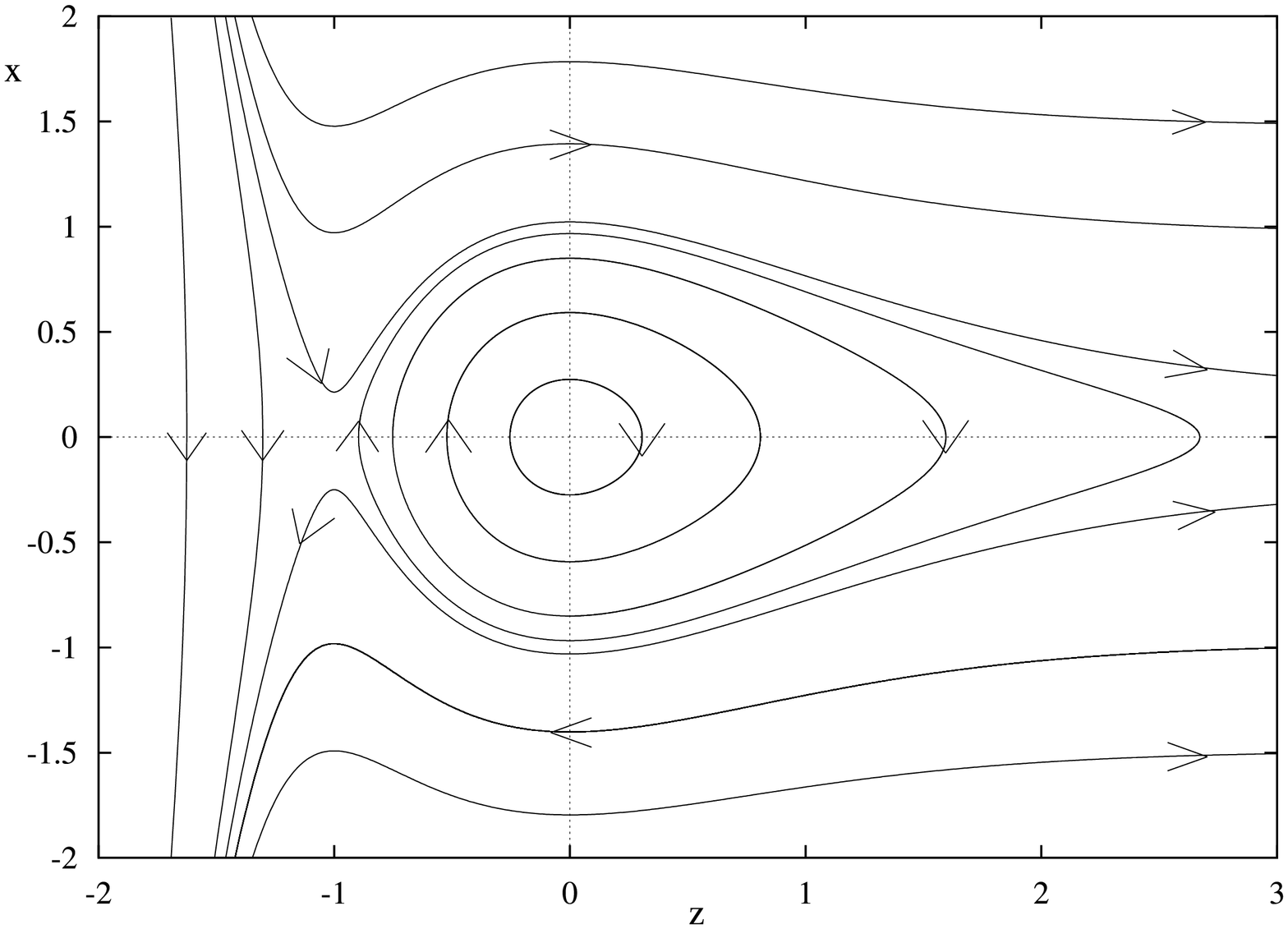}
\caption{The phase portrait $(z,x)$ for case {\bf B} $p = (\gamma_{0} + \gamma_{1}z)\rho$
($k = -1$, $\gamma_{0}=-1/3$, $\gamma_{1}=2/3$)}
\label{B3}
\end{center}
\end{figure}

For the case {\bf B} $\frac{dw}{dz} = \gamma_{1}$, the above condition means that $\gamma_{1}k > 0$.
The phase portraits for $\gamma_{0} = -1/3$, and various $\gamma_{1}$, are presented on Fig. \ref{B},
\ref{B1},\ref{B2} and \ref{B3}.

Let us note that critical points are located at
\begin{equation}
z_{0} = -\Bigg(\frac{1}{3 \gamma_{1}} + \frac{\gamma_{0}}{\gamma_{1}}\Bigg).
\label{qccrpt}
\end{equation}
If we put, for example, $\gamma_{0} = -1$ (cosmological constant term) then $\gamma_{1} > -\frac{3}{2}$ 
critical points can only exist on a finite domain of the phase plane because $z > -1$. 

The idea of structural stability comes from Andronov and Pontriagin \cite{andronow}. A~dynamical 
systems $S$ are said to be structurally stable if their dynamical behaviour remains qualitatively 
(modulo homeomorphizm preserving orientation of trajectories) the same (equivalent) under small perturbations.
Structural stability is sometimes considered as a precondition for the ``real existence''.  
Structurally stable dynamical systems on the 2D compact space (for example on the plane with circle at 
infinity) form open and dense subsets in the space of all dynamical systems on the plane. Therefore, 2D 
structurally unstable models seem to be nonadequate for describing real proceses because of measurement 
errors.

The main aim of the qualitative analysis of differential equations is not to find, and then to analyze, 
individual solutions but rather to investigate space of all possible solutions for all 
admissible initial conditions. A property is believed to be ``realistic'' if it can be attributed to large
(rather typical than exceptional) subsets of models within the space of all possible solutions, or if it
possesses a certain stability, i.e. if it is shared by a slighty perturbed model. There is a wide opinion 
among specialists that realistic models should be structurally stable, or even stronger, that everything
that exists should possess a kind of structural stability.

From the physical point of view it is interesting to answer the question: are the trajectories for which
acceleration of the universe takes place, distributed in a typical or exceptional way? How are trajectories
with interesting properties distributed in the phase plane? For example, the acceleration condition 
$\ddot{a} = - \partial V/ \partial a > 0$ is satisfied if $V(a)$ is a decreasing function of $a$.
For us it is important that one should be easily able to deduce this from the geometry of the potential 
function only. 

In the phase space, the area of acceleration is determined by the condition that
\begin{equation}
\frac{\ddot{a}}{a} = -\frac{1}{6}(1+3w)\rho > 0,\hspace{5mm} 
\frac{\partial V}{\partial a} + \psi(a)(a')^{2} < 0.
\label{accond1}
\end{equation}
(\ref{accond1}) can be rewritten to the form which could be usefull in the analysis of the  ``probability 
of acceleration'' which can be defined as a measure of the space of those initial conditions that lead 
to accelerating universes
\begin{align}
&-\frac{\partial}{\partial a} \ln{(E-V)}+\frac{1+3w(a)}{a} < 0, \nonumber \\
\intertext{or}
&(1+z)^{2}\frac{\partial}{\partial z}\ln{(E-V(z))}+(1+3w(a(z))(1+z) < 0. 
\label{accond2}
\end{align}

For example, domains of acceleration in the configurational space, for the case {\bf D}, as well as the 
domain of $\ddot{a} > 0$ ($\frac{\partial V}{\partial a} < 0$) in the phase plane are presented on 
Fig.~\ref{wyk10}.

Let us now consider the presence or absence of the particle horizon in the past. Good news from our earlier 
discussion is that this property can be detected from the shape of the diagram of the potential function of 
the system, namelly if $V(z)$ goes to a constant (zero is included) as $z \rightarrow \infty$ then we obtain
a model without the horizon. 

In the next section it will be demonstrated how we can answer the question about the horizon in the past on 
the base of $V(z)$ taken from the observations.

\section{Inverse problem in Quintessential Cosmology}

The presented formalism gives us a natural base to discuss the redshift magnitude relation $m(z)$ for SNIa
supernovae observational data. But on the other hand, because the Hubble function is related to the luminosity
distance, it is possible to determine both the quintessence parameter $w(z) = p/\rho$ and the potential
of the dynamical system $V(z)$. Therefore, the equation of state as well as the whole dynamics can be 
reconstructed provided that the luminosity function $d_{L}(z)$ is known from observations. It is called
the inverse problem in dynamics of quintessential cosmology.

As an example of constructing observables from the considered dynamics let us consider the 
luminosity--distance relation $d_{L}(z)$ for quintessential models. 

If a light source of redshift $z$ is located at a radial coordinate $r_{1}$ (taken from R--W metric), its
luminosity distance $d_{L}$, its angular diameter distance $d_{A}$ and its proper motion distance are given by
\begin{equation}
d_{L}(z) = (1+z)a_{0}r_{1},\hspace{5mm} d_{A} = \frac{a_{0}r_{1}}{1+z},\hspace{5mm} d_{M} = a_{0}r_{1},
\label{invlumdist1}
\end{equation}
where $r_{1}$ calculated from metric gives
\begin{equation}
\begin{split}
\varphi(r_{1}) &= \int^{a_{0}}_{\frac{a_{0}}{1+z}}\frac{da}{a\dot{a}} = 
\frac{1}{a_{0}}\int_{0}^{z}\frac{dz'}{H(z')} \\  
&= \begin{cases} \sin^{-1}{r_{1}} & \textrm{when } k=+1 \\
r_{1} & \textrm{when } k=0 \\
\sinh{r_{1}} & \textrm{when } k=-1 \end{cases} 
\end{split}
\label{invrad1}
\end{equation}
Here $a_{0}$ is the present value of the radius of the universe. The above equation can also be written in
the form of a single compact equation as
\begin{equation}
\frac{d_{L}(z)}{1+z} = \frac{1}{\sqrt{\kappa}}\zeta \Bigg(\sqrt{\kappa}\int^{z}_{0}\frac{dz'}{H(z')}\Bigg),
\label{invlumdist2}
\end{equation}
where
$$
\begin{array}{lll}
\zeta(q) = \sin{q} & \textrm{with } \kappa=\Omega_{k,0} & \textrm{when } \Omega_{k,0}>0, \\
\zeta(q) = \sinh{q} & \textrm{with } \kappa=-\Omega_{k,0} &\textrm{when }\Omega_{k,0}<0, \\
\zeta(q) = q & \textrm{with } \kappa=1 &\textrm{when }\Omega_{k,0}=0.
\end{array}
$$
Thus for a given $\mathcal{M}$ (absolute magnitude) and $H(z)$ equation (\ref{invlumdist2}) gives the 
predicted value of $m(z)$ (observed magnitude) at a given $z$.

By using the $\kappa$--corrected effective magnitudes $m_{i}$ which have also been corrected for the 
light curve width--luminosity relation and the galactic extinction, and by using the same standard errors 
$\sigma_{2,i}$ and $\sigma_{m_{i}}^{{\rm eff}}$ of the supernova with redshift $z_{i}$ as used by 
Perlmutter et al. we compute $\chi^{2}$ according to
\begin{equation}
\chi^{2} = \sum_{i}\frac{[m_{i}^{{\rm eff}}-m(z_{i})]^{2}}{\sigma_{z_{i}}^{2}+\sigma_{m_{i}\ {\rm eff}}^{2}}.
\label{invhikw}
\end{equation}
The best fit parameters are obtained by minimizing equation (\ref{invhikw}).

Luminosity distance and angular distance depend sensitively on the present densities of various energy 
components and their equations of state. For this reason, the magnitude--redshift relation for distant 
standard candles has been proposed as a potential test for cosmological models.

In our formalism $H(z)$ can be immediately taken from the first integral of the dynamical equation and then 
we obtain
\begin{equation}
\frac{d_{L}(z)}{1+z} = \frac{1}{\sqrt{\kappa}}\zeta \Bigg(\sqrt{\kappa}\int^{z}_{0}
\frac{e^{-3/2\int\frac{w(a(z'))}{1+z'}dz'}dz}{(1+z)^{3/2}\sqrt{2(E-V(z))}}\Bigg).
\label{invlumdist3}
\end{equation}
Formula (\ref{invlumdist3}) limits the determination of the luminosity distance because it depends on 
quintessential parameter $w(z)$ through a multiple integral relation that smears out detailed information
about $w(z)$. If $w(a(z))$ can be expanded as the Taylor series following (\ref{wrozw}) then we obtain the 
simplest formula without the double integration
\begin{equation}
\frac{d_{L}(z)}{1+z} = \frac{1}{\sqrt{\kappa}}\zeta \Bigg(\sqrt{\kappa}\int^{z}_{0}
\frac{\exp{\{-\frac{3}{2}\sum \gamma_{i}z'^{i} dz'\}}}{\sqrt{2(1+z')^{3(1+\gamma_{0})}(E-V(z'))}}\Bigg).
\label{invlumdist4}
\end{equation}
Many authors \cite{saini} assume that a quite accurate luminosity distance may be obtained and then examined
to answer the question of whether the equation of state of the expanding universe can be determined uniquely.

Our idea is more general. To determine the structure and evolution of an astrophysical system of the universe,
the equation of state is usually necessary. By equation of state of the universe we mean the relation 
between the total energy density of cosmic matter and the total pressure. However, the equation of state
relevant to the universe has not yet been established. Our idea is to reconstruct it from the form of the 
potential of the system $V(a(z))$.

Let us consider, for simplicity, a flat model for which the Hubble parameter is related to luminosity 
distance by the relation
\begin{equation}
H(z) = \Bigg[\frac{d}{dz}\Bigg(\frac{d_{L}(z)}{1+z}\Bigg)\Bigg]^{-1}.
\label{invhubbl}
\end{equation}
Then it is possible to determine the quintessence parameter
\begin{equation}
w(z) = -1 -\frac{2}{3}H(1+z)\frac{d^{2}}{dz^{2}}\Bigg(\frac{d_{L}(z)}{1+z}\Bigg).
\label{invquintess1}
\end{equation}
Here the term $-1$ is established from the condition that for stationary solution $H(z) = {\rm const}.$ 
$\frac{d^{2}}{dz^{2}}(\frac{d_{L}(z)}{1+z})$ will vanish but such a solution can appear on the phase plane 
$(H,\rho)$ only as an intersection trajectory of the flat model and the boundary of the weak energy 
condition $\rho+p=0$.

\begin{figure}[!ht]
\begin{center}
\includegraphics[scale=0.5]{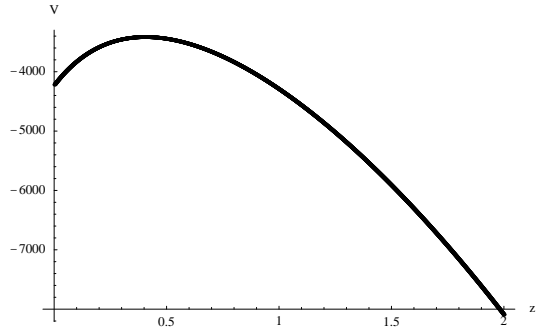}
\caption{The plot of the potential function $V(z)$ has been obtained from equation (\ref{invpot}) and from 
fitting formula (\ref{fitfun}) to the SNIa observational data}
\label{snpot}
\end{center}
\end{figure}

Equation (\ref{invquintess1}) can be rewritten in the form
\begin{align}
p &= - \rho - 3\xi(z)H, \\
\intertext{where}
\xi(z) &= \frac{2}{3}H^{2}(z)(1+z)\frac{d^{2}}{dz^{2}}\Bigg(\frac{d_{L}(z)}{1+z}\Bigg). \nonumber
\label{invquintess2}
\end{align}
The dependence $p(H)$ manifests the presence of bulk viscosity effects in the model 
$\xi(z) = -\frac{1}{3}\frac{\partial p}{\partial H}$. Therefore, from $d_{L}(z)$ we obtain the reconstruction
of $w(z)$ as a mixture of the cosmological constant term and bulk viscosity. However, in determining $w(z)$
there is an inherent limitation because the value $w(z)$ is poorly resolved and no usefull constraint can be 
obtained concerning its time variation. Of course, the value of $w(0)$ and $dw/dz$ with accuracy $0.1$
and $0.15$ respectively will be obtained from SNAP3 but for the reconstruction of $w(z)$ higher derivatives 
can be useful. If we measure $\frac{dw^{i}}{dz}(0)$, $i \geqslant 2$, then in principle, the reconstruction 
equation of the equation of state is possible, because then
$$w(z) = w(0) +\frac{dw}{dz}(0)z + \frac{1}{2}\frac{d^{2}w}{dz^{2}}(0)z^{2} + \ldots,$$
where $\frac{d^{n}w}{dz^{n}}(0)$ is determined from the recurrence formula, namelly
\begin{equation}
\begin{split}
\frac{d^{n}w}{dz^{n}}(0) &  =  \frac{2}{3}\frac{d^{n}}{dz^{n}}\Bigg|_{0}(\ln{H}) + \frac{2}{3}(1+z)\frac{d^{n+1}}{dz^{n+1}}\Bigg|_{0}(\ln{H}), \\
w(0) & = - 1 + \frac{2}{3}\frac{d}{dz}\Bigg|_{0}(\ln{H}).
\end{split}
\label{invquintess3}
\end{equation}
Then $w(z)$ can be obtained from $H(z)$ in the following way
\begin{equation}
\begin{split}
H(z) & \mapsto  \ln{H(z)} \mapsto \forall n \frac{d^{n+1}}{dz^{n+1}}(\ln{H}) \\
& \mapsto  \frac{d^{n}w}{dz^{n}}(0) \mapsto w(z) = \sum_{i/1}^{\infty}\frac{1}{i!}\frac{d^{i}w}{dz^{i}}(0)z^{i} + w(0).
\end{split}
\label{invquintess4}
\end{equation}
At present this idea can not be realized but the idea of reconstructing dynamics requires the knowledge of 
$V(z)$ which can be calculated from the Hamiltonian constraint 
\begin{equation}
\begin{split}
V(a(z)) &\equiv - \frac{\rho_{{\rm eff}}a^{2}(z)}{6} = - \frac{1}{2} H^{2}(z)a^{2}(z) \\
&= - \frac{1}{2} \Bigg[\frac{1}{(1+z)\frac{d}{dz}\big(\frac{d_{L}(z)}{1+z}\big)}\Bigg]^{2}.
\end{split}
\label{invpot}
\end{equation}
On the other hand, the knowledge of $V(z)$ gives us information how the horizon problem can be solved. 
Reconstruction of the potential function from SNIa data is presented on Fig.\ref{snpot}. We obtain the plot
of the potential function from fitting
\begin{equation}
\begin{split}
\frac{d_{L}(z)}{1+z} &= \int_{0}^{z}\Bigg[A_{0}+A_{1}(1+z')+A_{2}(1+z')^{2}+{} \\
&+ A_{3}(1+z')^{3}+A_{4}(1+z')^{4}\Bigg]^{-\frac{1}{2}}dz'
\end{split}
\label{fitfun}
\end{equation}
function to the SNIa observational data.

We can observe that as $z \rightarrow \infty$, $V(z)$ as calculated from (\ref{invpot}) goes to $-\infty$. 
This can be treated as an empirical evidence of the presence horizon in the past. Let us note that other 
problems of the standard cosmology can be discussed analogously basing on information about $V(z)$ and its 
geometry.

\section{Conclusions}

In this work a class of FRW models with quintessence matter is examined in the context of the present 
acceleration of the universe. Our results are the following.

\begin{enumerate}
\item
We have given a mathemathical background for discussing physical content of quintessential cosmology. 
Its dynamics is reduced to the dynamics of the unit mass point particle in 1D potential. Then different 
physical properties, like acceleration of the universe, existence of horizon, can be formulated only in terms
of the potential function of the system. The proof of the corresponding condition is quite general. In 
particular, it is independent of any specific assumption about the behaviour of the scale factor near the 
singularity (such as the assumption of power low behaviour) or a specific form of equation of state.
\item
The dynamics is formulated in the hamiltonian formalism and the full classification of possible evolutions 
in the phase plane as well as in configurational space is given. In the near future, it will be possible to
obtain, from SNAP3, the exact value of $\gamma_{0}$ and $\gamma_{1}$ appearing in the equation of state 
$p = (\gamma_{0} + \gamma_{1}z)\rho$, and then we could automatically answer the question about the
horizon.
\item
The effectiveness of our treatment of dynamics of quintessential models in terms of single-particle mechanics
is demonstrated for a broad class of tracking potentials. We obtain topological equivalence of the phase 
portraits (for this case) with the dynamical system obtained for the potential function reconstructed from 
SNIa data.
\item
The idea of reconstructing dynamics instead of quintessential coefficient is considered. It is what we called
inverse problem in quintessential cosmology. We demonstrate that the reconstructed potential function of the
system produces the particle horizon in the past and solves the flatness problem.
\end{enumerate}

\begin{acknowledgments}
Authors are very greatefull to prof M. Heller and dr W. Godlowski for discussion and comments.
\end{acknowledgments}

\onecolumngrid

\newpage

\begin{figure}[!hp]
\begin{center}
\includegraphics[scale=0.3, angle=270]{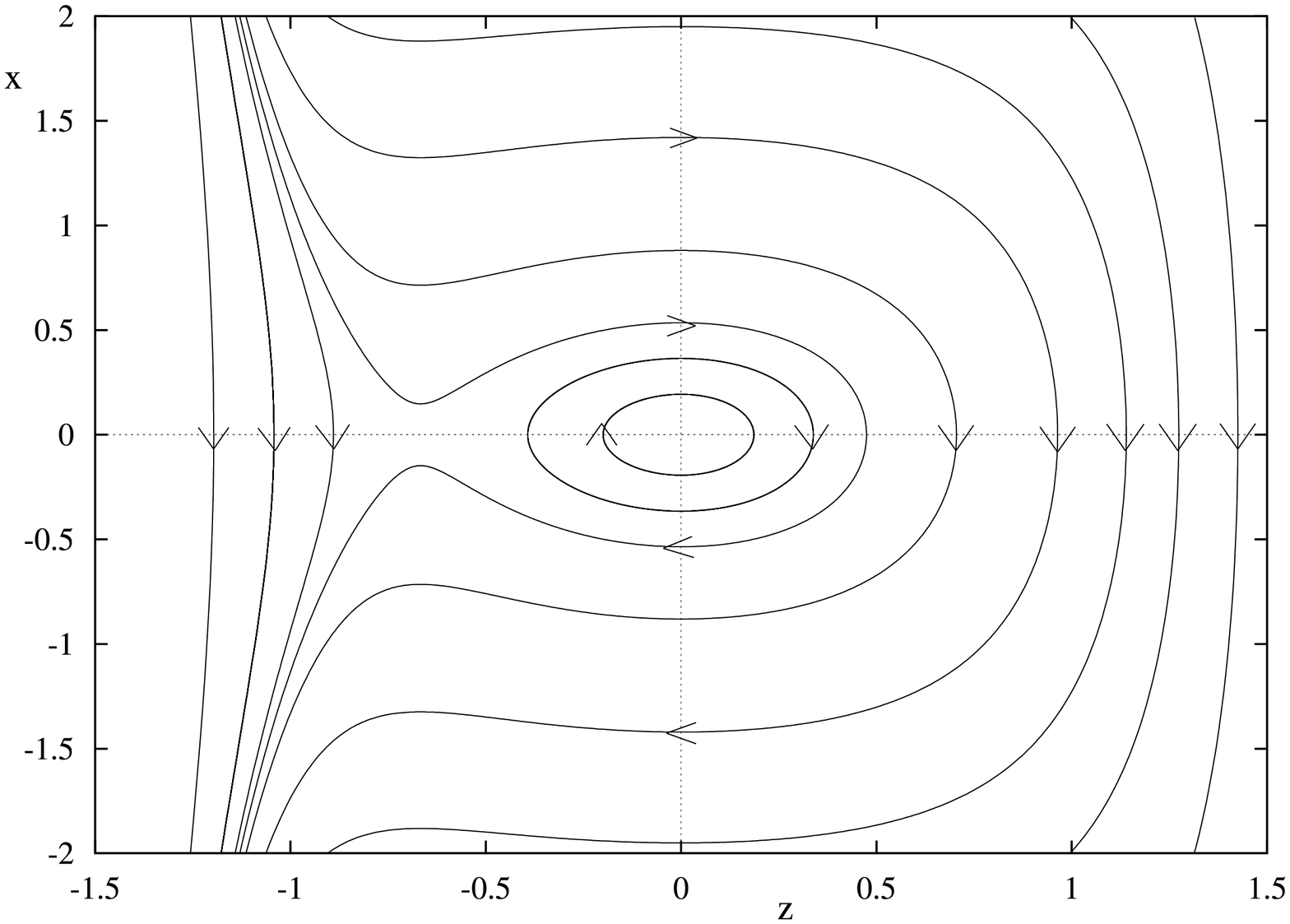}
\caption{The phase portrait $(z,x)$ for case {\bf A} $p = (\gamma_{0} + \gamma_{1}z + \gamma_{2}z^{2})\rho$ 
($k = 1$, $\gamma_{2} = -1$)}
\label{A}
\end{center}
\end{figure}

\begin{figure}[!hp]
\begin{center}
\includegraphics[scale=0.3, angle=270]{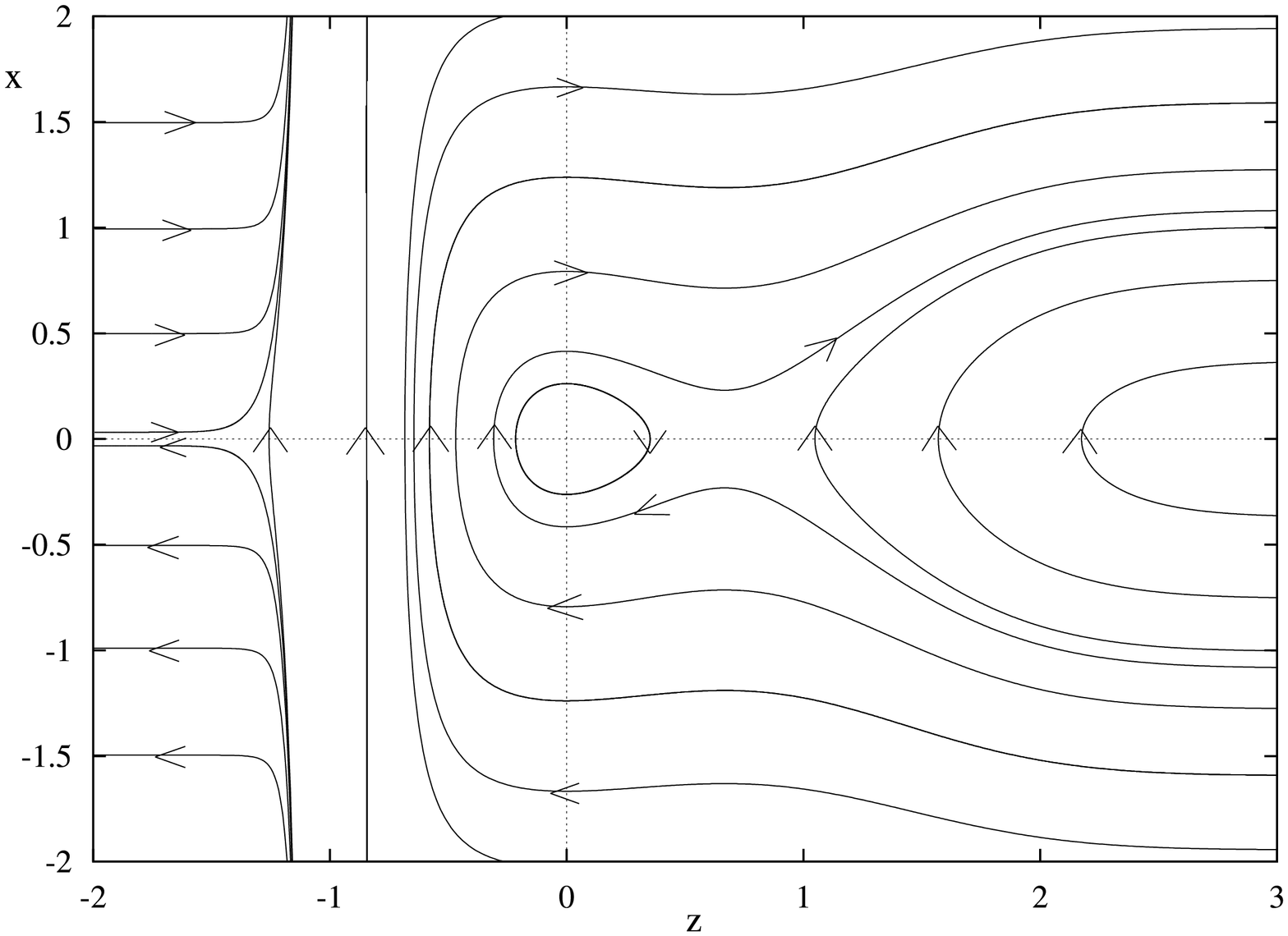}
\caption{The phase portrait $(z,x)$ for case {\bf A} $p = (\gamma_{0} + \gamma_{1}z + \gamma_{2}z^{2})\rho$ 
($k = 1$, $\gamma_{2} = 1$)}
\label{A1}
\end{center}
\end{figure}

\begin{figure}[!hp]
\begin{center}
\includegraphics[scale=0.3, angle=270]{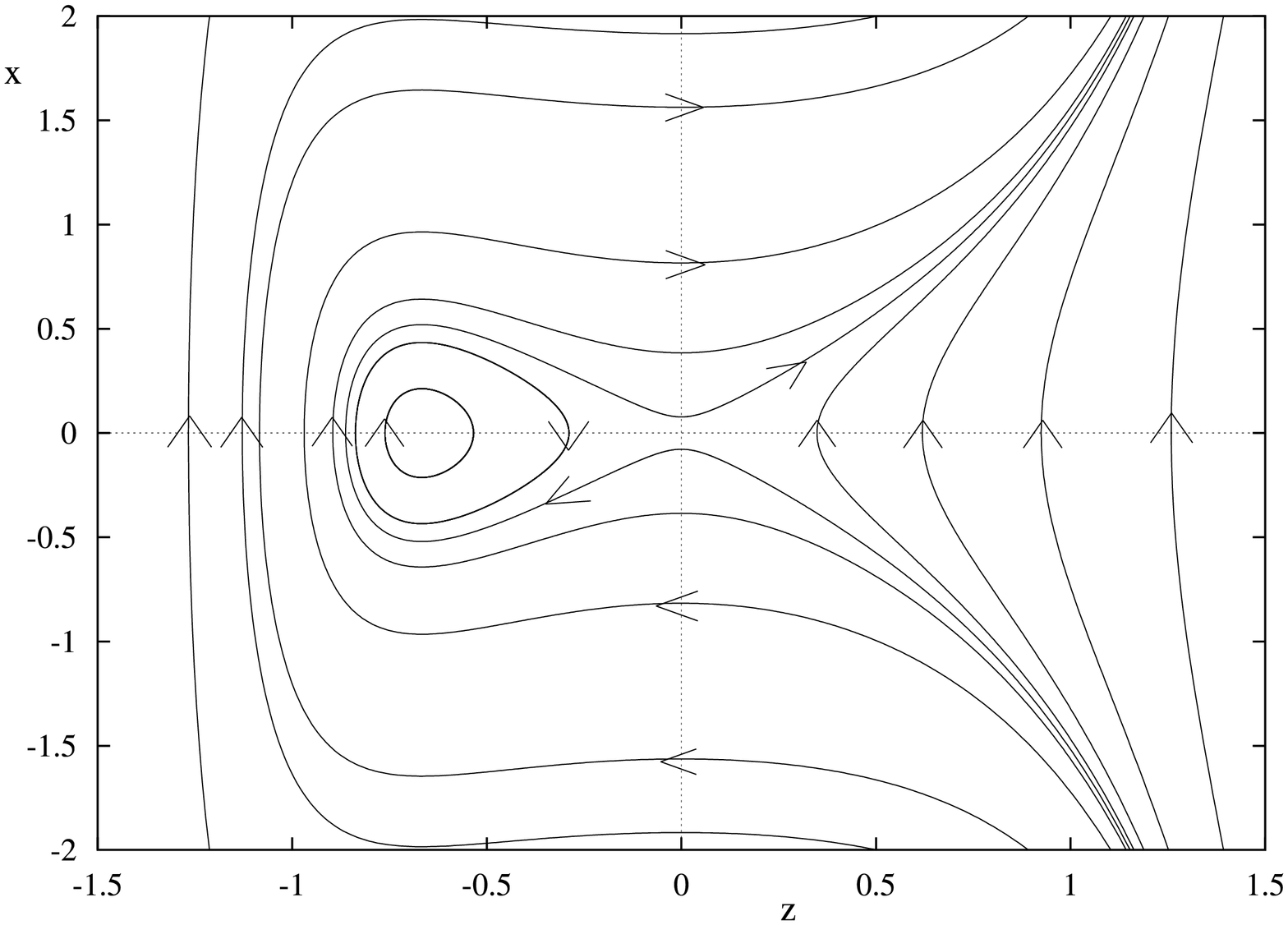}
\caption{The phase portrait $(z,x)$ for case {\bf A} $p = (\gamma_{0} + \gamma_{1}z + \gamma_{2}z^{2})\rho$
($k = -1$, $\gamma_{2} = -1$)}
\label{A2}
\end{center}
\end{figure}

\begin{figure}[!hp]
\begin{center}
\includegraphics[scale=0.3, angle=270]{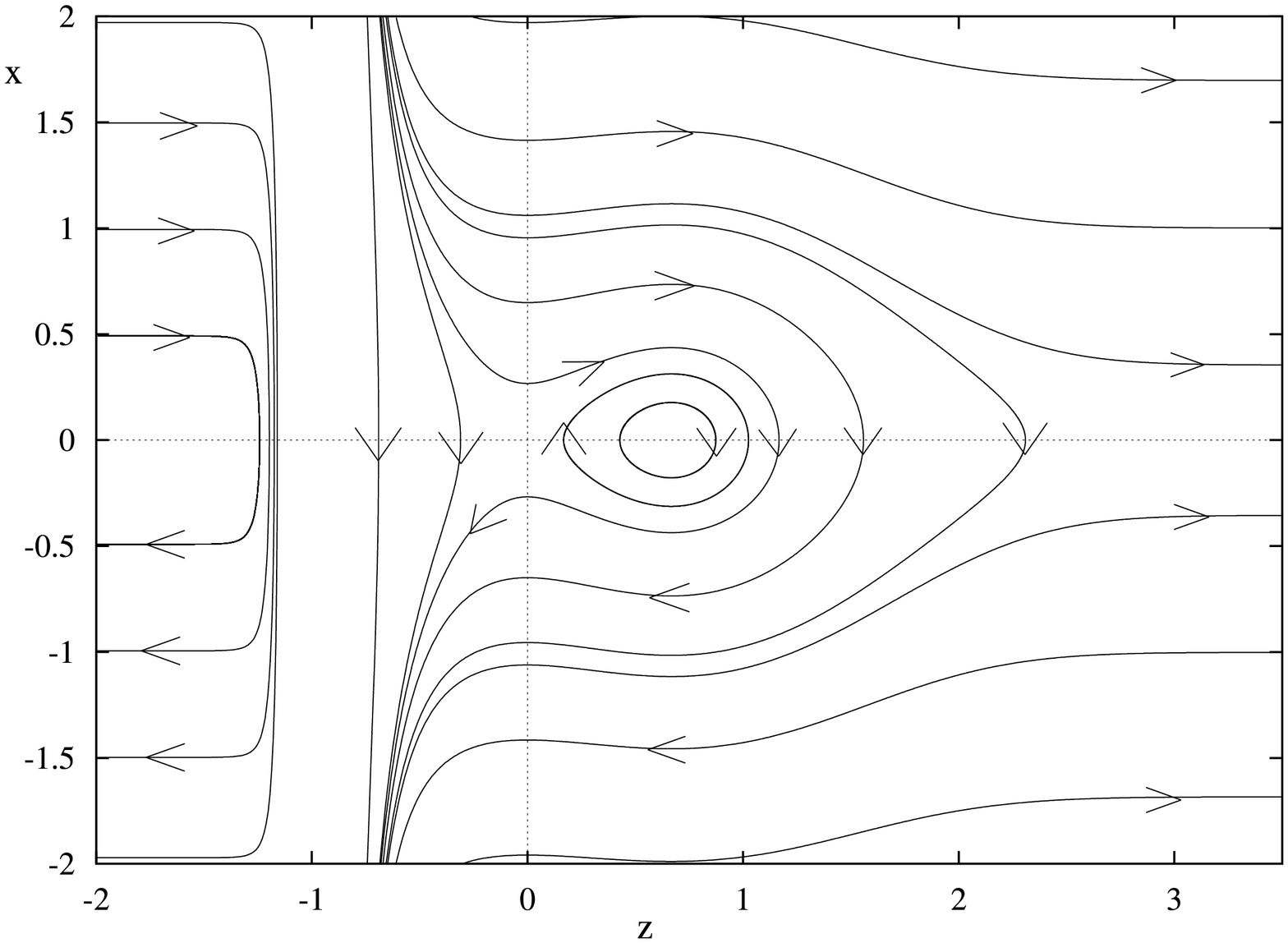}
\caption{The phase portrait $(z,x)$ for case {\bf A} $p = (\gamma_{0} + \gamma_{1}z + \gamma_{2}z^{2})\rho$
($k = -1$, $\gamma_{2} = 1$)}
\label{A3}
\end{center}
\end{figure}

\begin{figure}[!hp]
\begin{center}
\includegraphics[scale=0.3, angle=270]{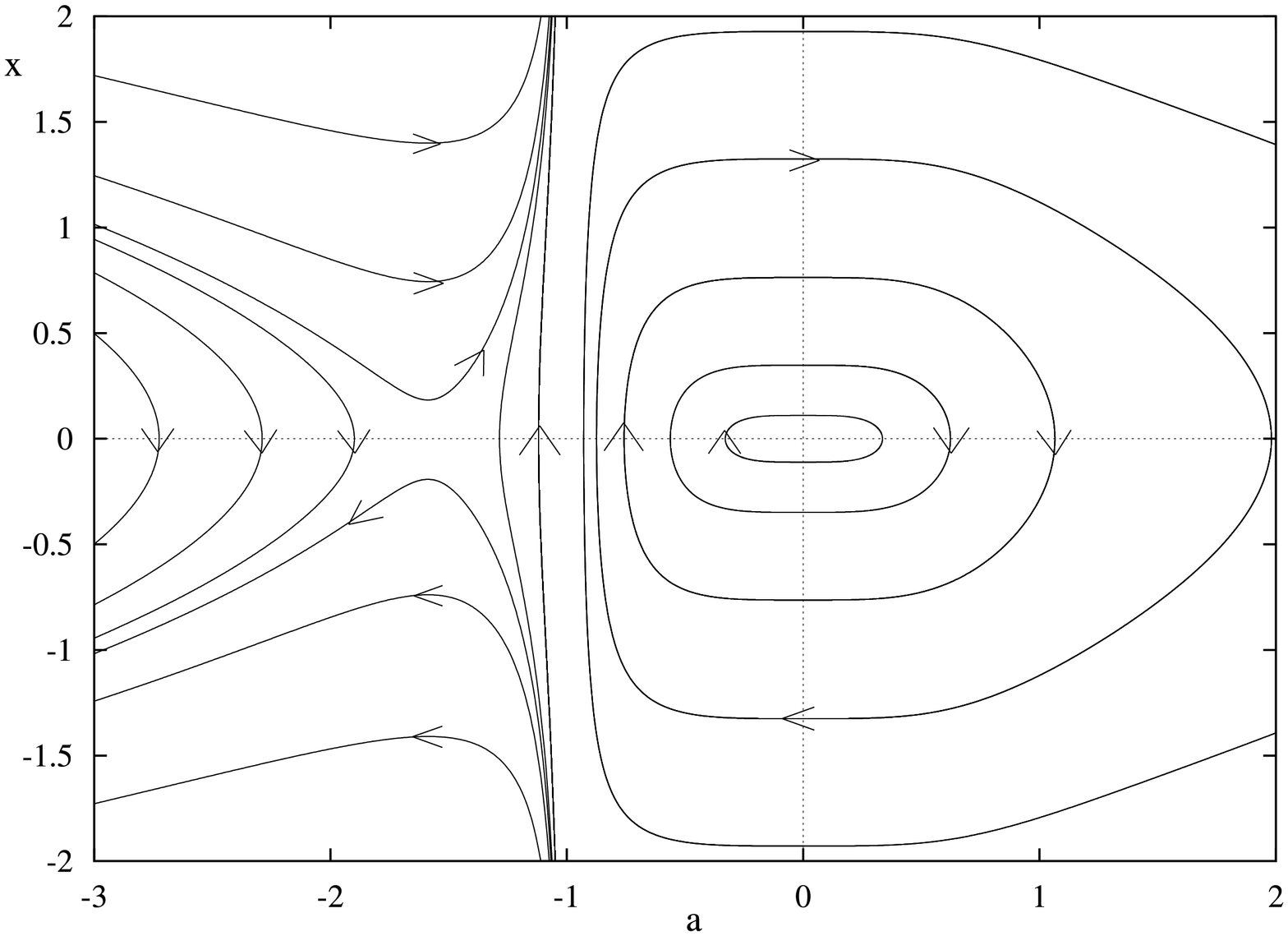}
\caption{The phase portrait $(a,x)$ for case {\bf C} $p = w_{x}\rho_{x}$, 
$k = 1$, $\rho_{0m}/\rho_{0x} = 1$, $w_{x} = 1$ }
\label{C}
\end{center}
\end{figure}

\begin{figure}[!hp]
\begin{center}
\includegraphics[scale=0.3, angle=270]{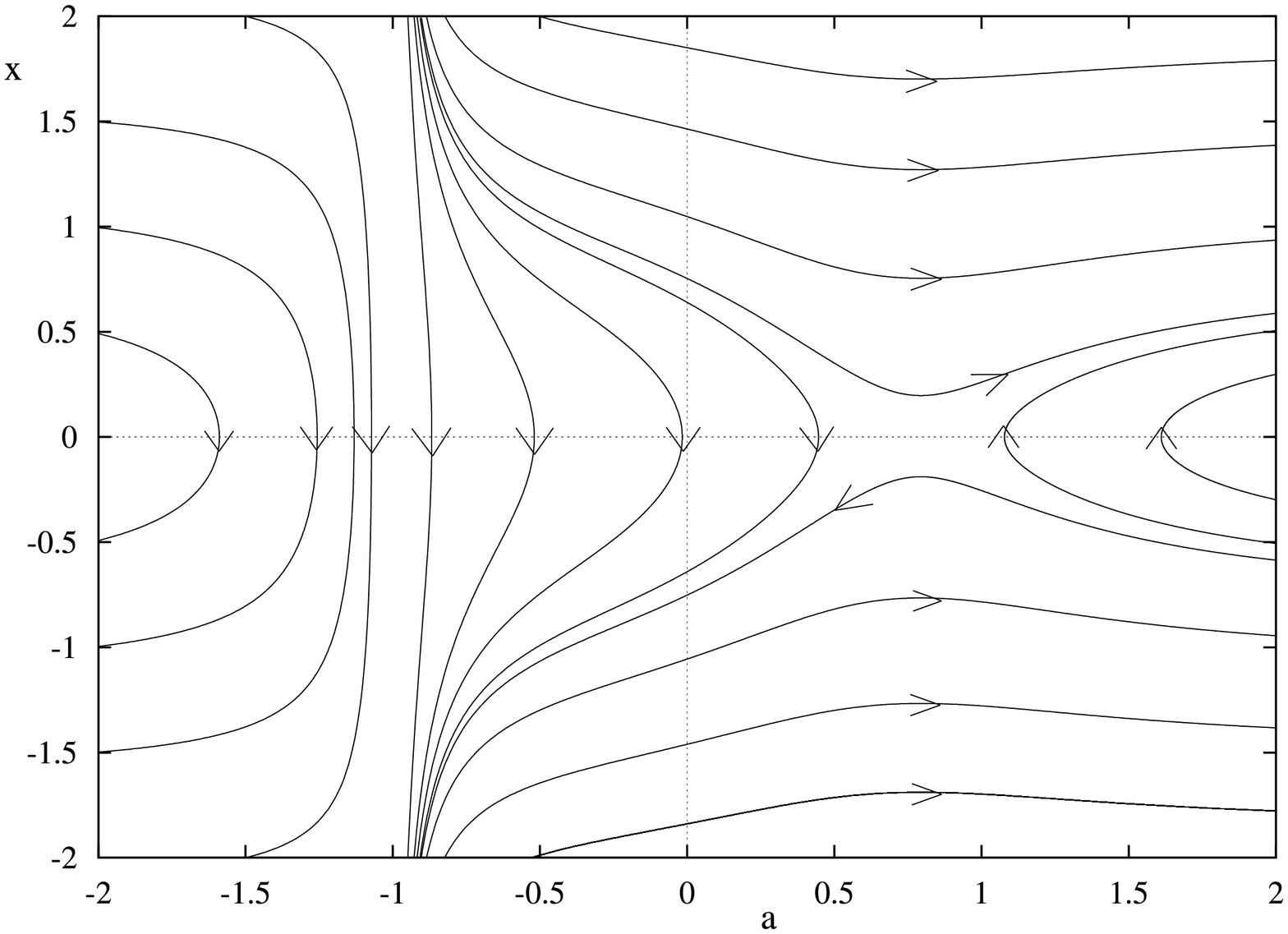}
\caption{The phase portrait $(a,x)$ for case {\bf C} $p = w_{x}\rho_{x}$,
$k = 1$, $\rho_{0m}/\rho_{0x} = 1$, $w_{x} = -1$}
\label{C1}
\end{center}
\end{figure}

\begin{figure}[!hp]
\begin{center}
\includegraphics[scale=0.3, angle=270]{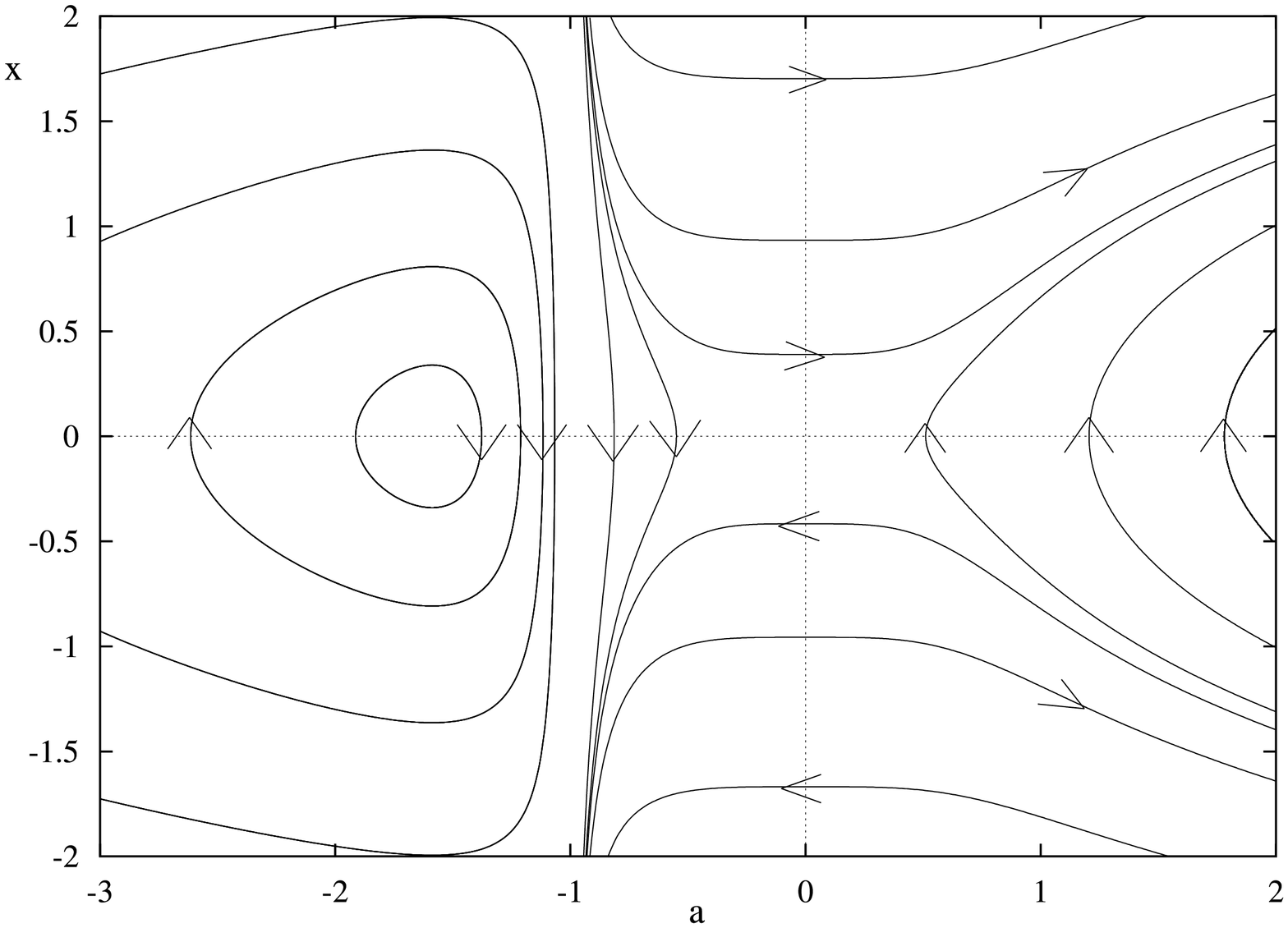}
\caption{The phase portrait $(a,x)$ for case {\bf C} $p = w_{x}\rho_{x}$,
$k = -1$, $\rho_{0m}/\rho_{0x} = 1$, $w_{x} = 1$ }
\label{C2}
\end{center}
\end{figure}

\begin{figure}[!hp]
\begin{center}
\includegraphics[scale=0.3, angle=270]{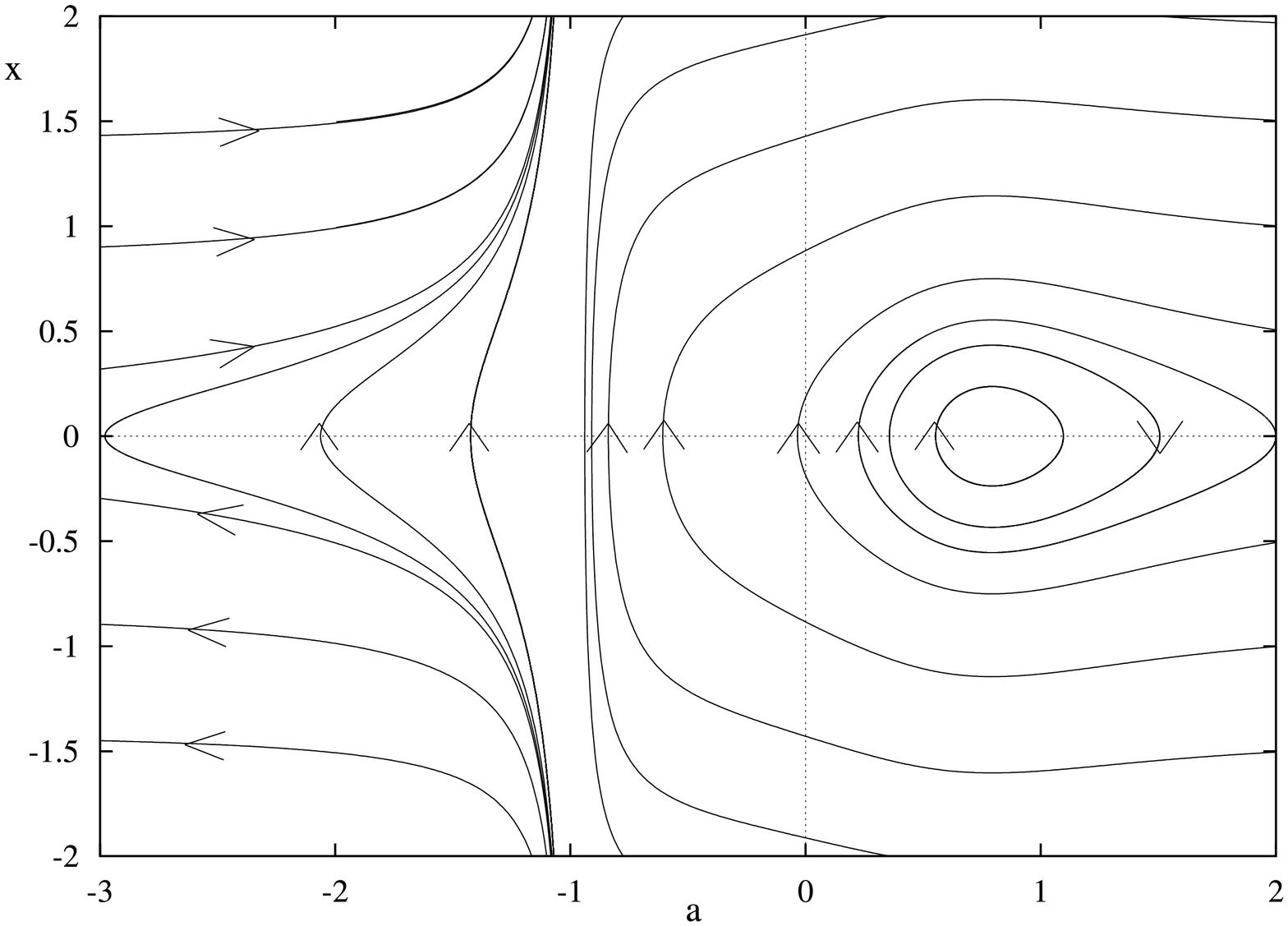}
\caption{The phase portrait $(a,x)$ for case {\bf C} $p = w_{x}\rho_{x}$,
$k = -1$, $\rho_{0m}/\rho_{0x} = 1$, $w_{x} = -1$}
\label{C3}
\end{center}
\end{figure}

\end{document}